\documentclass[fleqn,10pt]{article}
\usepackage[utf8]{inputenc}
\usepackage[T1]{fontenc}

\usepackage[margin=1in]{geometry}
\usepackage{hyperref}
\usepackage{soul}
\usepackage{subcaption}
\usepackage{graphicx}%
\usepackage{multirow}%
\usepackage{array}
\usepackage{amsmath,amssymb,amsfonts}%
\usepackage{amsthm}%
\usepackage{mathrsfs}%
\usepackage[title]{appendix}%
\usepackage{xcolor}%
\usepackage{textcomp}%
\usepackage{manyfoot}%
\usepackage{booktabs}%
\usepackage{algorithm}%
\usepackage{algorithmicx}%
\usepackage{algpseudocode}%
\usepackage{listings}%
\usepackage{rotating}%
\usepackage{lineno}%
\usepackage{accents}
\usepackage{titling}

\title{Time-Resolved Multi-Spectral X-ray Computed Tomography of Cryoprotectant Diffusion Into Biomimetic Material}

\author{Alaa M. Ali$^+$,\\
        \and
        Jason T. Parker$^+$,\\
        \and
        Anthony N. Consiglio,\\
        \and
        Brooke S. Chang,\\
        \and
        Linnea Warburton,\\
        \and
        Boris Rubinsky,\\
        \and
        Simo A. M\"akiharju}

\begin{document}

\date{}
\flushbottom
\maketitle
\begin{center}
    Department of Mechanical Engineering, University of California, Berkeley, Berkeley, CA 94720, USA
    
    $^+$These authors contributed equally to this work.
\end{center}

\thispagestyle{empty}

\section*{Abstract}

Cryopreservation via vitrification requires loading cryoprotective cocktails. Insufficient loading may lead to freezing, precluding successful recovery; over-loading is toxic. Yet, existing in-situ measurements of cryoprotectant permeation remain largely unvalidated and do not resolve individual cryoprotectant concentrations. We introduce multi-spectral X-ray computed tomography (MSCT) to non-invasively quantify the spatiotemporal distribution of cryoprotectants diffusing into a tissue-mimicking phantom. A developed photon-energy bin selection algorithm achieves sensitivity to low-contrast cryoprotectants without contrast agents or fluorescence edges. The technique is validated with a dimethyl sulfoxide–glycerol–water solution, resolving cryoprotectant volume fractions to within 5\% accuracy. We observe heterogeneous diffusion of the cryoprotectants into the tissue-mimicking hydrogel, a phenomenon not observable with conventional techniques. MSCT improves upon existing X-ray CT methods because it is not underdetermined for multi-component solutions and does not implicitly assume homogeneous diffusion. These advancements enable the systematic development of cryoprotectant loading protocols and provide diagnostics to assess vitrifiability before cryopreservation.

\section*{Introduction}

Simultaneous diffusive transport of multi-component mixtures is a widespread phenomenon in biological, industrial, and environmental systems \cite{lohse_fundamental_2022, lohse_physicochemical_2020, hyman_liquid-liquid_2014}. Spatiotemporally resolved measurements of individual components are needed to fully characterize multi-component transport systems. However, measuring diffusion in multi-component fluid systems is challenging due to fluid opacity, refractivity, and sensitivity to invasive probes.

X-ray image-based techniques are advantageous for fluid component decomposition because they overcome these limitations. First, X-rays can non-invasively pass through many optically opaque systems. Second, X-rays have an index of refraction near unity, overcoming refractivity challenges at interfaces. Lastly, X-ray attenuation is a function of the photon energy and material composition. Many methods have been developed to take advantage of these properties for material decomposition. For example, for materials with fluorescence edges (particularly K-edges) in an achievable energy range, one can use multi-energy CT with switched source voltages, filters, dual-layer detectors, or a combination thereof \cite{russo_handbook_2018, garnett_comprehensive_2020, granton_implementation_2008}. However, many materials of medical and industrial importance -- especially low molecular weight solutes -- lack fluorescence edges in the \textit{O}(10-100~keV) range typical of laboratory and medical X-ray systems, thus limiting the applicability of K-edge-based approaches for those materials.

Multi-spectral CT (MSCT) is useful in cases without K-edges. MSCT acquires X-ray CT images at multiple discrete photon energies to calculate the composition of the material. The scan(s) can be captured by either tuning a monochromatic source (e.g., at a synchrotron) or with energy-resolved photon counting X-ray detectors that bin detected photons by energy. In principle, MSCT can decompose an arbitrary number of materials provided CT images are taken with at least as many photon energies as there are materials (although in practice this may not be achievable for reasons discussed later). Prior research on the MSCT technique has focused on medical applications for enhancing image contrast and quantifying contrast agent concentration \cite{curtis_quantification_2019, brendel_empirical_2009, sato_multi-energy_2024}. The focus on image contrast enhancement and contrast agents is due in part to limitations on the materials that can be decomposed. MSCT depends on X-ray attenuation contrast between the constituent materials for the decomposition step. Low molecular weight materials frequently found in biomedical applications often have low X-ray attenuation contrast across the range of photon energies typical of laboratory and medical X-ray sources, making decomposition difficult without contrast agents.

An application where these challenges are particularly acute and which would benefit from non-invasive, spatiotemporally resolved composition measurements is vitrification-based cryopreservation. Successful vitrification of biological samples allows for their stable long-term storage and, when applied to complex samples, has the potential to substantially improve entire fields. For example, the field of medical transplantation would be greatly improved by the successful vitrification and storage of organs, enabling increased availability of healthy organs and the ability to predictably schedule organ transplant surgeries. Vitrifying tissues and organs requires loading cryoprotective anti-freeze cocktails by diffusive uptake or vascular perfusion and rapidly cooling the sample to achieve a glassy solid phase, avoiding the formation of crystalline ice, which can damage sensitive tissue structures \cite{powell-palm_cryopreservation_2023, han_vitrification_2023, namsrai_cryopreservation_2025}. To achieve ice-free vitrification (and rewarming recovery) in bulk tissues, aqueous solutions containing cryoprotective agents (CPAs) must be introduced into the tissue to replace much of the freezable water and reduce the critical cooling and warming rates from \textit{O}$(10^6)$\textdegree C/min for pure water \cite{bruggeller_complete_1980} to rates feasible in bulk tissues -- \textit{O}$(1)$\textdegree C/min.  Ensuring complete permeation of CPAs into the specimen is imperative for successful vitrification to avoid local ice formation and is the focus of perfusion protocol development \cite{brockmann_normothermic_2009, corral_optimized_2018, hessheimer_normothermic_2019, leonel_stepped_2019}. Perfusion protocols must balance the fundamental engineering trade-off between the need for complete permeation of CPAs into the tissue and the cost of toxicity damage from the CPAs.  Ideally, the sample is exposed to the lowest feasible concentration of CPAs for the shortest viable amount of time while achieving a high enough concentration of CPAs within the tissue to avoid ice upon cooling and rewarming.

At present, these vitrification protocols are typically developed with limited information on the CPA distribution within the specimen. During CPA perfusion in an organ, for example, the concentration of the venous effluent is measured using methods such as refractometry to determine whether a threshold concentration has been reached, yet a singular measurement cannot provide information on the distribution of the CPA throughout a three-dimensional tissue. With such limited data, CPA formulations and loading/unloading protocols are iterated upon through trial and error, with protocols being evaluated for successful physical preservation via the detection of crystalline ice after cooling. The parameter space for CPA formulations, loading/unloading protocols, and toxicity limits is enormous, and progress towards successful vitrification is constrained by the limited ability to evaluate CPA distributions in situ. Diagnostic tools for assessing CPA loading, validation data sets for perfusion modeling, and a means to determine when CPA levels are just right or toxically high would significantly benefit the advancement of vitrification.

The ice detection step to determine successful vitrification is often performed visually, but X-ray computed tomography (CT) has also been commonly used to detect ice formation in vitrified specimens \cite{bischof_use_2007} and CPA solutions \cite{parker_direct_2023, ali_experimental_2024}. X-ray CT has also been regularly used to measure CPA concentration in biosamples \cite{corral_optimized_2018, bischof_use_2007, han_diffusion_2020, corral_assessment_2015, corral_use_2021, manuchehrabadi_improved_2017}. However, as we will show, standard energy-integrating CT measurements, while useful, are limited to requiring either binary CPA solutions (e.g., DMSO and water) or that all components of a CPA solution diffuse homogeneously. Most modern CPA solutions contain several components, and there is evidence that different molecules permeate at different rates \cite{higgins_permeation_2023, warner_osmotic_2023, ahmadkhani_high_2025}. Some CPA components can, for example, penetrate cell membranes while others cannot \cite{bleisinger_me2so_2020, jomha_permeation_2009}, and solutions containing combinations of these CPAs would be expected to permeate into biomaterials heterogeneously. These measurements have not been performed in multi-component solutions.

In this work, we introduce a laboratory-scale MSCT technique together with a practical energy-bin selection algorithm to non-invasively decompose multi-component cryoprotectant cocktails into their individual components. The energy bin optimization analysis is designed to be simple to implement in a practical laboratory setting without extensive imaging system characterization. We first validate the method on a prototypical cryopreservation solution of known composition containing water, dimethyl sulfoxide (DMSO), and glycerol. Next, applying it to quantitatively characterize the diffusion of this solution into a tissue-mimicking hydrogel phantom, we directly observe the heterogeneous diffusion of DMSO and glycerol predicted by previous studies yet unresolved by conventional energy-integrating CT techniques. We conclude by discussing the implications of this new measurement technique for rational, image-guided design of vitrification protocols and investigating simultaneous transport in multi-component fluid systems in general.

\section*{Results}

\subsection*{Overview of Measurement Apparatus}

\begin{figure}
    \centering
    \includegraphics[width=0.85\linewidth]{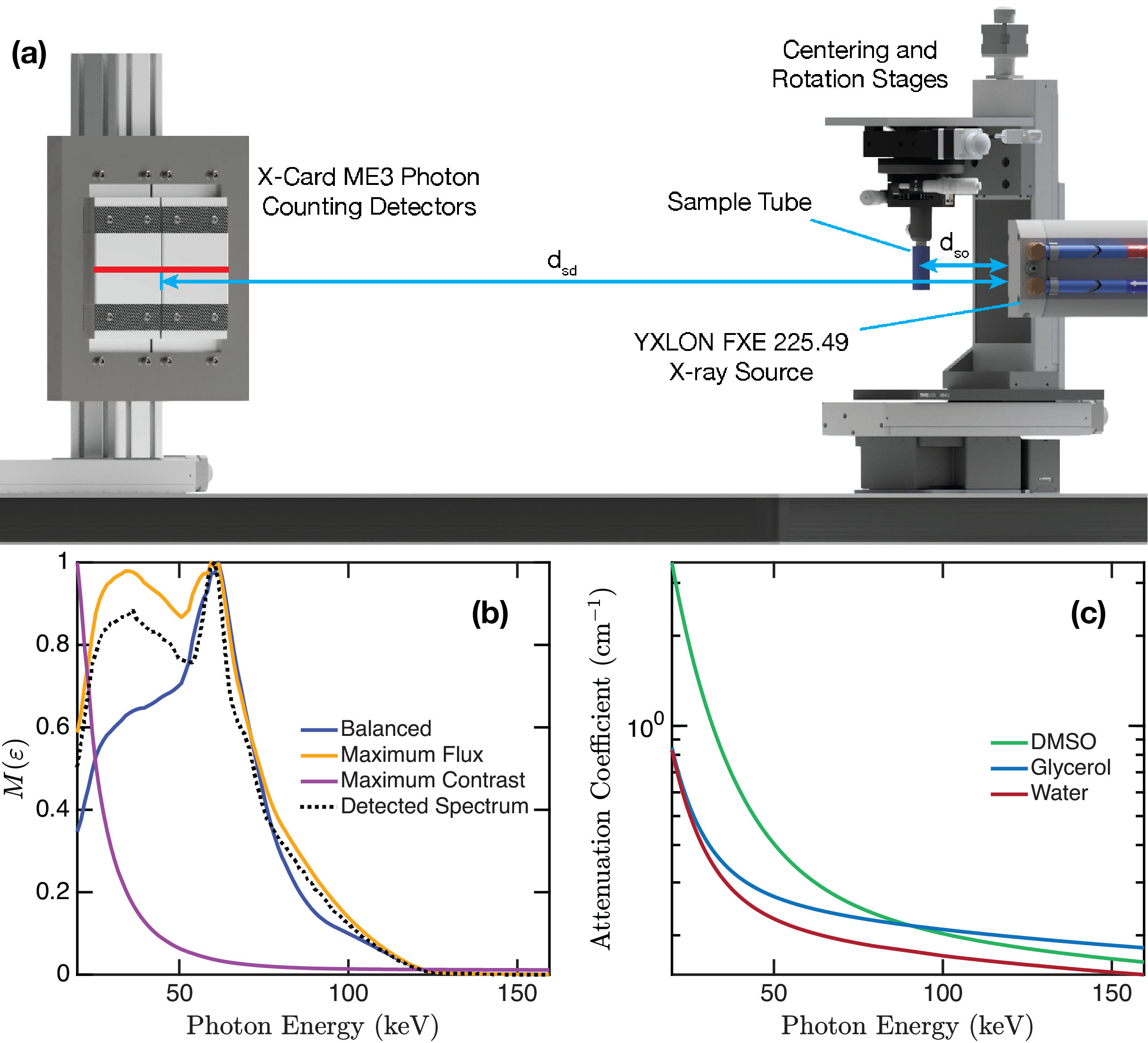}
    \caption{(a) The X-ray source is a tungsten-target YXLON FXE225.99 X-ray source with the directional head installed. The source emits a polychromatic cone beam with a 15\textdegree~angle. The detector in this study is a Detection Technologies X-Card ME3 line detector capable of up to 128 photon energy bins. The approximate line detection region is indicated with a red line on the detector. (Not shown in figure \ref*{fig:setup} is a custom cooling system to keep the ME3 detector cool during operation.) The sample is in a 15~mL conical tube, which is connected from above to a stack of rotation and centering stages. The source to detector distance, $d_{sd}$, is 1,367~mm; the source to object distance, $d_{so}$, is 140~mm. (b) The measured spectrum and the energy metric, $M(\varepsilon)$, for the range of emitted photon energies. Note that the balanced strategy does not necessarily choose the bins with the highest flux. (c) The linear attenuation coefficients for the three liquid components: DMSO, glycerol, and water. At low photon energies the DMSO has great contrast with glycerol and water, but glycerol and water have low contrast with one another. The balanced energy bin contrast metric considers this in choosing the optimal bins while the maximum contrast strategy does not.}
    \label{fig:setup}
\end{figure}

MSCT as implemented here requires a polychromatic X-ray source and an energy-resolving photon counting detector (PCD). We use a custom X-ray CT system, shown in Figure \ref*{fig:setup}a, to capture scans. The X-ray source is an YXLON FXE 225.99 with the directional head. A combination of two ME3 energy-resolved photon counting detectors takes line images. The ME3 detector can use up to 128 energy bins, enabling multi-spectral imaging in different energy bins simultaneously. Such a detector has the advantage of enabling higher temporal resolution imaging because no source-, detector-, or filter-switching is required. Validation solutions or tissue-mimicking hydrogel samples in solution are placed in a 15~mL conical tube attached to the centering and rotation stages.

\subsection*{MSCT Decomposition and Bin Selection Algorithm}

In standard X-ray CT, assuming that photon energy-integrated variables adhere to the Beer-Lambert law, the reconstruction intensity can be decomposed into a function of the solute concentration. Equation \ref*{eq:ctdecomp} shows how the reconstruction intensity $\tilde{R}(\mathbf{x})$ can be decomposed to calculate the combined solutes concentration, $c_U(\mathbf{x})$.
\begin{equation} \label{eq:ctdecomp}
    \tilde{R}(\mathbf{x}) = \tilde{\varsigma}_U(\mathbf{x}) c_{U}(\mathbf{x}) + \tilde{\varsigma}_V c_V= A(\mathbf{x}) c_U(\mathbf{x}) + B
\end{equation}
Here, $c_{V}$ is the solvent concentration; $\tilde{\varsigma}$ is the energy-integrated mass attenuation coefficient; $\tilde{\cdot}$ indicates a variable pertains to energy-integrated CT. Energy-integrated X-ray CT can only be decomposed into two components, otherwise the system is under-determined. A calibration is done to determine the coefficients $A(\mathbf{x})$ and $B$. In the case of a binary solution $A(\mathbf{x})$ can be considered constant since the solute mass attenuation is a material property. In multi-component systems, however, constant $A(\mathbf{x})$ requires that the combined solutes mass attenuation coefficient, $\tilde{\varsigma}_U(\mathbf{x})$ is constant. In turn, this implies that the components must diffuse homogeneously, as shown by equation \ref*{eq:constatten} (see Supplemental Information for the derivation):
\begin{equation} \label{eq:constatten}
    \varsigma_U(\mathbf{x}) = \frac{ \sum_{i \in U} \mu_i \phi'_i(\mathbf{x}) }{ \sum_{j \in U} \rho_j \phi'_j(\mathbf{x}) } = \text{constant}.
\end{equation}
Here, $\varsigma_{U}$ is the combined mass attenuation coefficient for the set of solutes, $U$; $\mu_i$ is the linear attenuation coefficient of component $i$; $\phi'_i \equiv v_i / v_U$ is the volume fraction of component $i$ with respect to the solutes (not the overall solution). For $\varsigma_U(\mathbf{x})$ to be constant $\phi'_i(\mathbf{x})$ must be constant for all $i$. Since by construction the standard X-ray CT method only decomposes into two components -- solvent and solute -- it is impossible to know if that condition is met. So, it is often implicitly assumed to be met. Strictly speaking, in standard energy-integrated X-ray CT the energy-integrated mass attenuation, $\tilde{\varsigma}_U$, must be constant. In equation \ref*{eq:constatten} we use the regular mass attenuation since we later calculate $\phi'_i$ from energy-resolved experiments.

In MSCT, no such implicit assumptions are required. The reconstruction voxel intensity is the combined linear attenuation coefficient of the materials in that voxel. Assuming that the density in solution is roughly equal to the density of the material, that combined linear attenuation coefficient is a linear combination of the component linear attenuation coefficients and their respective volume fractions, as shown in equation \ref*{eq:decomprecon} for $K$ components.
\begin{equation}\label{eq:decomprecon}
    R(\mathbf{x}, E_i) = R_i(\mathbf{x}) = \sum_{j=1}^K \mu_{j}(E_i)\phi_j(\mathbf{x}) = A_{ij} \phi_j(\mathbf{x})
\end{equation}
Here, $R(\mathbf{x}, E_i)$ is the reconstruction from images in energy bin $E_i$, $\phi_j$ is the volume fraction of component $j$ and $\mu_j$ is the linear attenuation coefficient of material $j$. In addition to equation \ref*{eq:decomprecon} there are the constraints that the volume fractions by definition must sum to unity and that they must all be between 0 and 1 \cite{yoon_image_2018},
\begin{align}
    \sum_{j=1}^K &\phi_j = 1 \\
    &\phi_j \in [0, 1]
\end{align}

Regularized least squares minimization, shown in equation \ref*{eq:lsqlin}, is used to calculate the volume fraction of each component in each voxel of the reconstruction. This decomposition step is performed separately inside the segmented hydrogel and outside the hydrogel. Outside the hydrogel we do not include the hydrogel component in the decomposition.
\begin{equation}\label{eq:lsqlin}
    \phi(\mathbf{x}) = \min_{\phi^*} \Big[ \Vert \mathbf{A} \phi^* - \mathbf{R}(
        \mathbf{x}) \Vert ^2_{2} + \lambda \Vert \mathbf{I} \phi^* \Vert^2_{2} \Big]
\end{equation}
Here, $\phi(\mathbf{x})$ is the volume fraction vector containing the volume fraction of each component at location $\mathbf{x}$, $\mathbf{R}(\mathbf{x})$ is the vector of reconstruction image intensities in each energy bin, $\lambda$ is the regularization parameter, and $\mathbf{I}$ is the identity matrix.

Tikhonov regularization \cite{golub_tikhonov_1999} is used due to the low contrast between water, glycerol, and hydrogel linear attenuation coefficients. The regularizing parameter $\lambda$ improves the numerical stability of the scheme and conditioning of the matrix $\mathbf{A}$. The derivation of equation \ref*{eq:decomprecon} and the implementation details of regularized least squares minimization and the choice of $\lambda$ are provided in the Supplemental Information.

\subsection*{Energy Bin Selection Algorithm Validation}

The choice of energy bins affects the decomposition quality. It becomes increasingly important to choose advantageous bins in systems involving components with low contrast attenuation coefficients. A novel energy bin selection algorithm (also referred to as the energy bin optimization) is developed that balances high detected photon counts with high collective X-ray attenuation contrast. We refer to this algorithm as the balanced strategy. Two other rational strategies are tested for comparison to assess the performance of the balanced energy bin selection strategy. The first is the maximum flux strategy, where the highest photon count bins are used. The second is the maximum contrast strategy, where bins with the highest overall X-ray contrast are chosen.

The energy bins are selected by choosing the bins that maximize the energy metric, $M(E_i)$, defined as
\begin{equation} \label{eq:binopt}
    M(E_i) \equiv \text{movmean}_{E_i}\left( C(E_i) \cdot I(E_i) \right) = \text{movmean}_{E_i}\left( \left\Vert \frac{  \Delta(E_i) }{ \left\Vert \Delta(E_i) \right\Vert_2 } - \mathbf{1}_S \right\Vert_2^{-1} \cdot I_{det}(E_i) \right)
\end{equation}
where $C(E_i)$ is the contrast metric as a function of energy bin $E_i$, $\Delta(E_i)$ is the vector of X-ray attenuation contrast between all combinations of two components, and $I(E_i)$ is the number of detected photons. $\mathbf{1}_S$ is the normalized unity vector in $\mathbb{R}^S$-space where $S \equiv \binom{K}{2}$ is the number of component combinations. Essentially, it is when all combinations of two components have the same contrast. The derivation of equation \ref*{eq:binopt} is detailed in the Supplemental Information. The energy metric is shown in figure \ref*{fig:setup}b for each bin selection strategy.

Maximum contrast and maximum flux use different contrast metrics than the balanced strategy. The former maximizes the overall attenuation contrast between all components regardless of flux; the latter maximizes the detected photon flux regardless of component attenuation contrast. In the maximum contrast strategy,
\begin{equation} \label{eq:maxcont}
    C(E_i) = \Vert \Delta(E_i) \Vert_2.
\end{equation}
High contrast between two components for a given energy can dominate the contrast metric even if the other components in the solution do not have high contrast for that same energy. Figure \ref*{fig:setup}c shows that at low photon energies the DMSO contrast with water and glycerol will dominate over the fact that there is low contrast between water and glycerol. Hence, in figure \ref*{fig:setup}b we see a high energy bin metric at low photon energies because DMSO contrast dominates. In the maximum flux strategy $C(E_i) = 1$, so only flux is considered. Note in figure \ref*{fig:setup}b that the balanced strategy does not always choose the bins with the highest detected photon counts. The resulting energy bins for all three methods are shown in table \ref*{tab:energybins}.

\begin{table}
    \centering
    \begin{tabular}{cccccc}
    \multicolumn{2}{c}{Balanced Set} & \multicolumn{2}{c}{Maximum Flux Set} & \multicolumn{2}{c}{Maximum Contrast Set} \\
    \hline
    Optimal Bins & Bins Used & Optimal Bins & Bins Used & Optimal Bins & Bins Used \\
    \hline
    \hline
    36.9 - 40.2 & 37.4 - 40.7 & 29.2 - 30.2 & 29.7 - 30.8 & 20.4 - 21.5 & 20.9 - 22.0 \\
    \hline
    42.4 - 45.7 & 42.9 - 46.2 & 32.5 - 35.8 & 33.0 - 36.3 & 24.7 - 25.9 & 25.3 - 26.4 \\
    \hline
    48.4 - 57.2 & 48.4 - 57.2 & 38.0 - 41.3 & 38.5 - 41.8 & 28.1 - 29.2 & 28.6 - 29.7 \\
    \hline
    58.9 - 64.4 & 59.4 - 64.9 & 55.0 - 63.8 & 55.0 - 63.8 & 31.4 - 32.5 & 31.9 - 33.0 \\
    \hline
    \end{tabular}
    \caption{The energy bins (keV) used for the different bin-choice strategies in ascending energy order (not ascending optimality order). The bins output by the program are in increments of 1.1~keV. The detector bins may not perfectly align with the bins output by the selection algorithm, which is the cause of the discrepancy between the optimal bins and the bins used.}
    \label{tab:energybins}
\end{table}

Due to the low X-ray attenuation contrast of the components the linear system of equations in (\ref*{eq:decomprecon}) has an ill-conditioned coefficient matrix which requires regularization to ensure stable solutions \cite{golub_tikhonov_1999}. Tikhonov regularized least squares minimization is used to decompose the component volume fraction. To control for the effect of regularization we compare the strategies with both equal regularization parameter or roughly equal condition number.

\begin{figure}
    \centering
    \begin{subfigure} [b] {0.45\textwidth}
        \centering\includegraphics[width=0.9\textwidth]{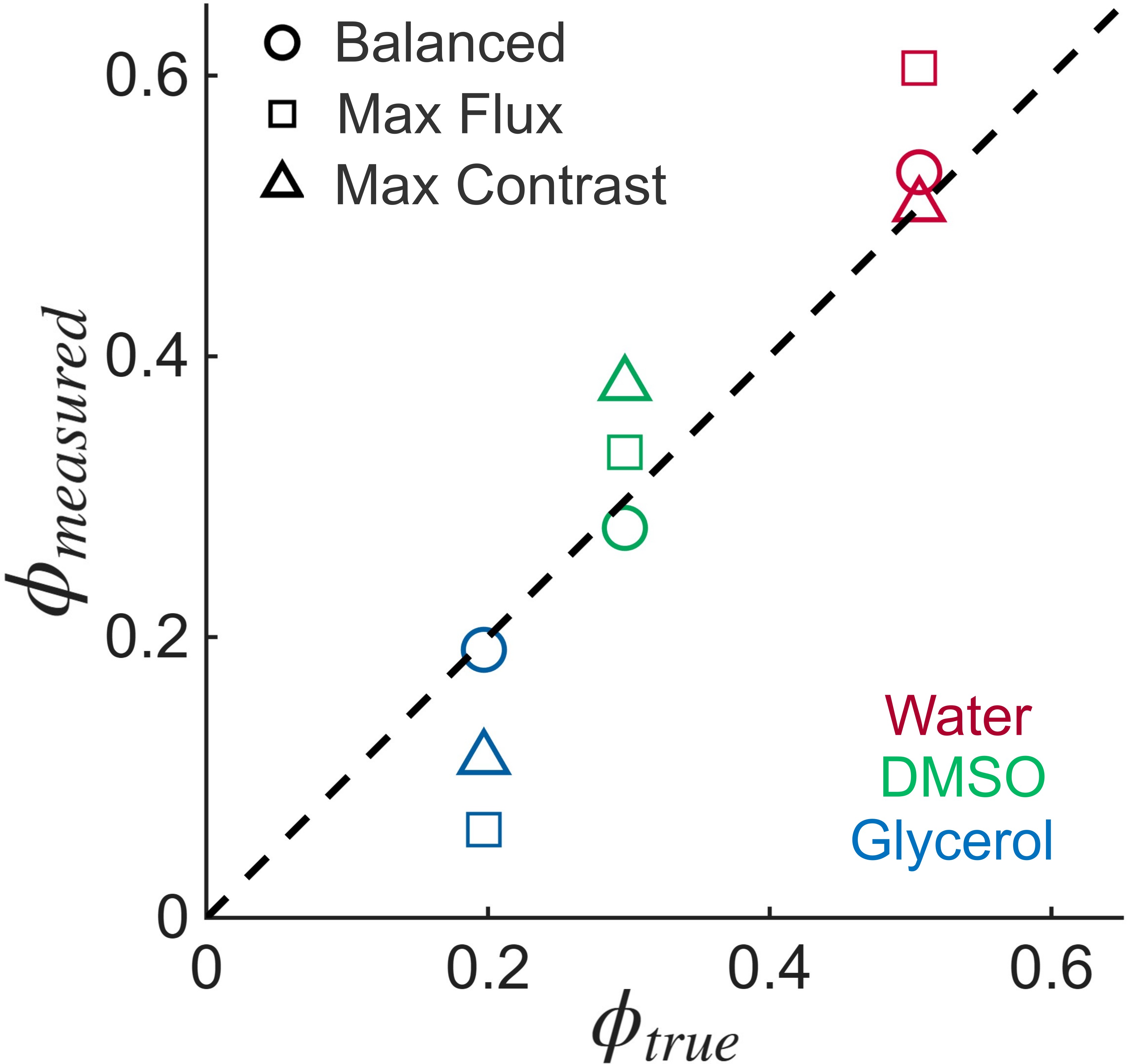}
        \caption{Solution 1}
        \label{fig:solution1_energies}
    \end{subfigure}
        \begin{subfigure} [b] {0.45\textwidth}
        \centering\includegraphics[width=0.9\textwidth]{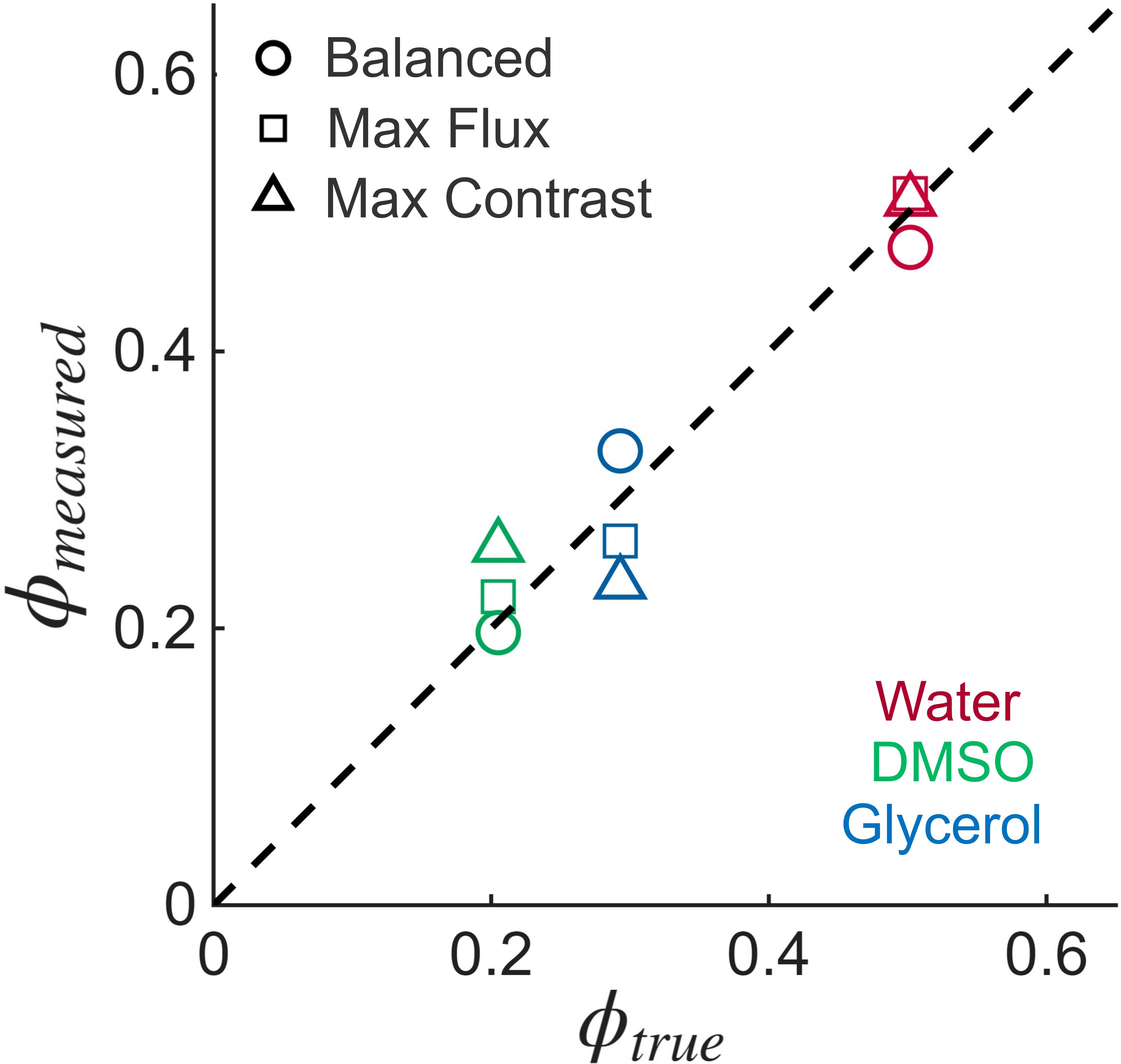}
        \caption{Solution 2}
        \label{fig:solution2_energies}
    \end{subfigure}
    \caption{The mean measured volume fractions are plotted against the known true volume fractions for the two validation solutions using the different bin selection strategies. The developed energy bin selection algorithm, the balanced strategy, outperforms the maximum flux and maximum contrast strategies for the same regularization parameter, $\lambda = 7.91 \times 10^{-4}$, across both solutions. The maximum flux strategy performs comparably to the balanced strategy in solution 2, but is outperformed in solution 1. The maximum contrast strategy is consistently outperformed. Especially in the case of solution 1 the maximum contrast strategy struggles to accurately decompose the glycerol and DMSO.}
    \label{fig:decomperror}
\end{figure}

Figure \ref*{fig:decomperror} shows the mean measured volume fraction against the true volume fraction for equal regularization parameter $\lambda = 7.91 \times 10^{-4}$. For both solutions the balanced bin strategy consistently outperforms the other two strategies. The maximum flux strategy is comparable to the balanced strategy in solution 2, but performs worse than the balanced strategy in solution 1. Even when the matrix condition numbers are matched at roughly 40, the balanced bin strategy outperforms the maximum flux and maximum contrast strategies. Table 4 in the Supplemental Information shows the mean error for each component in both solutions for the matched regularization parameter and condition number cases. The MSCT method can accurately measure the volume fraction of all components for both solutions when using the balanced bin selection strategy.

\subsection*{Validation on Ternary Solutions}

\begin{figure}
    \centering        
    \includegraphics[width=0.95\textwidth]{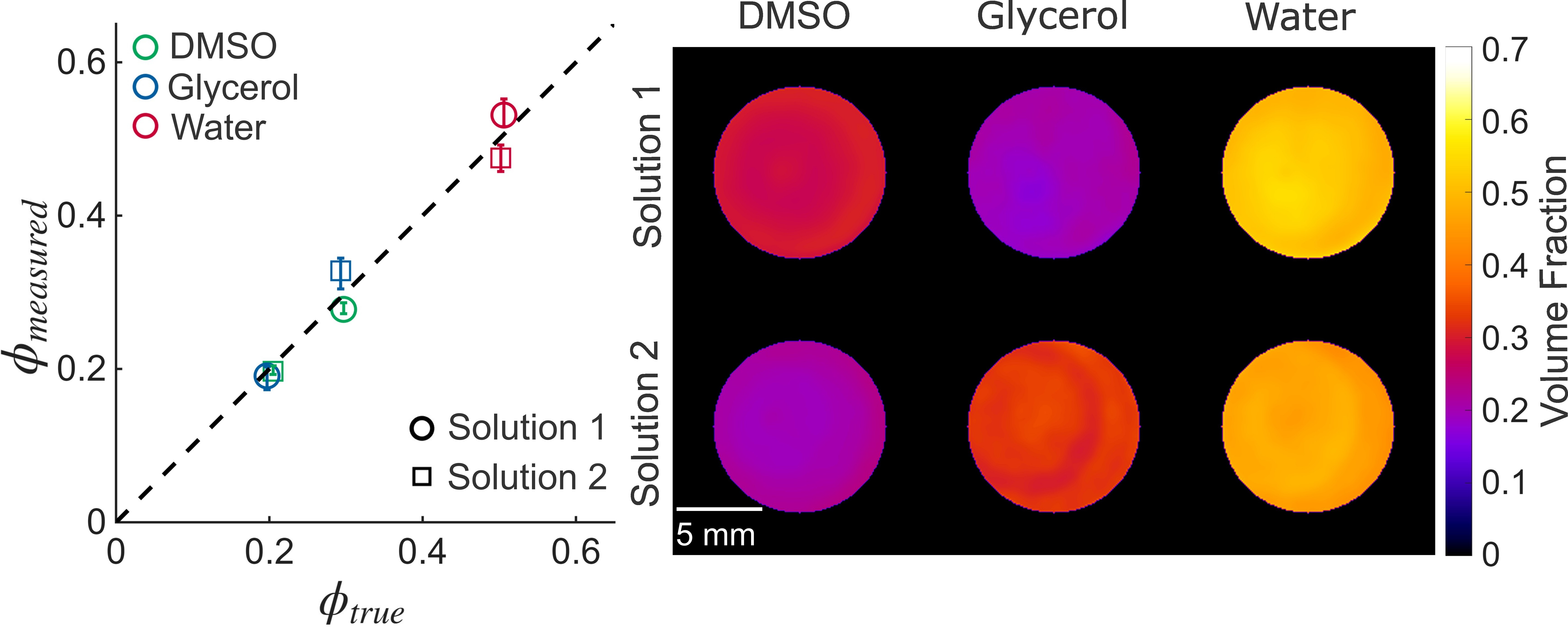}
    \caption{Volume fraction measurements of two well mixed solutions of known composition using the selected balanced energy bins compared to the known, true value. MSCT is able to measure the volume fraction accurately to within 5\% error. The error bars show the 5th and 95th percentiles of volume fraction values of each component within the decomposed reconstructions, showing the precision of the decomposition. No significant image artifacts are observed.}
    \label{fig:solndecomp}
\end{figure}

After selecting the energy bins, we validate the MSCT method by applying it to two well-mixed ternary solutions. Each solution has a different concentration of DMSO and glycerol. Figure \ref*{fig:solndecomp} shows the measured versus true volume fraction and decomposition cross sections of each of the two fluids. The mean error -- calculated as the absolute difference between the mean measured volume fractions and the known true volume fractions -- for DMSO, glycerol, and water is 1.9\%, 0.6\% and 2.5\% volume fraction, respectively, in the first solution and 0.8\%, 3.5\% and 2.7\% volume fraction in the second solution. The MSCT decomposition, then, shows good agreement with the known solution composition.

\subsection*{Hydrogel Diffusion}

\begin{figure}
    \centering
    \includegraphics[width=0.94\textwidth]{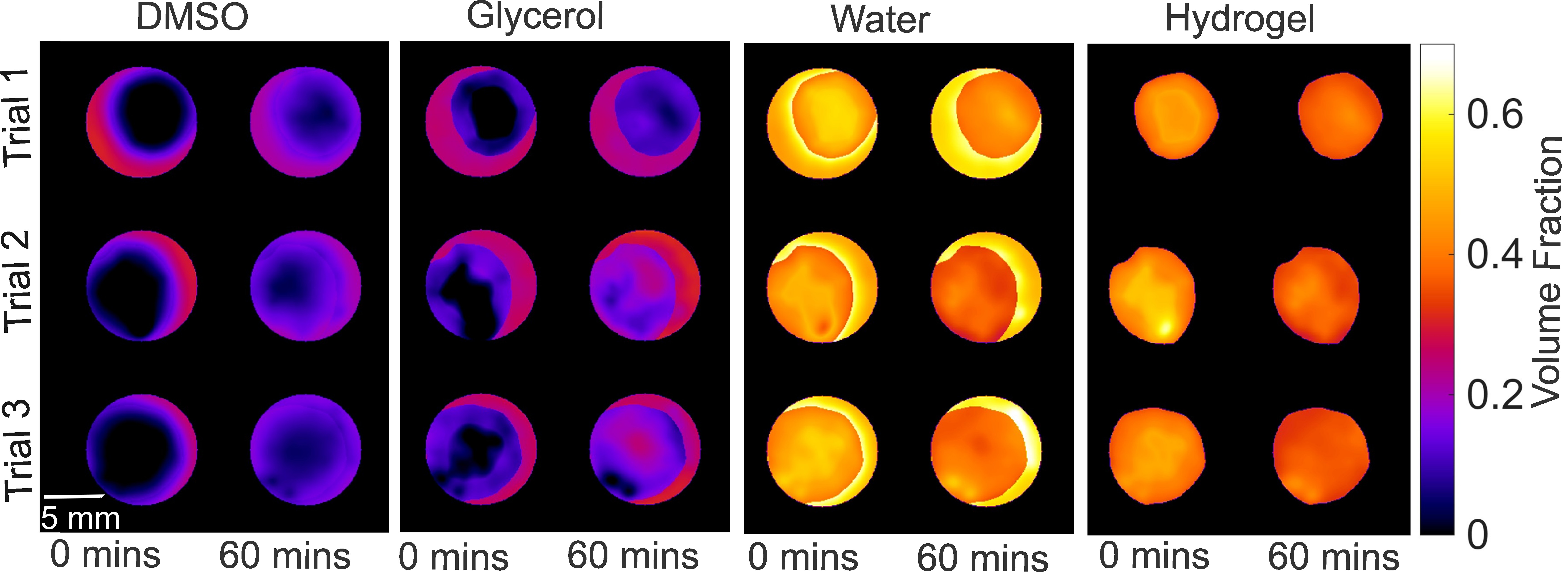}
    \caption{Volume fraction measurements of three representative trials immediately after inserting the hydrogel and one hour after inserting the hydrogel. The DMSO and glycerol are not observed to diffuse uniformly. Some diffusion is observed over the course of the 15~min of the first scan. Interestingly, glycerol is observed to diffuse more quickly than DMSO in this hydrogel. The volume fraction of water decreases as glycerol and DMSO diffuse into the hydrogel.}
    \label{fig:geldecomp}
\end{figure}

Figure \ref*{fig:geldecomp} shows the diffusion of each of the solution components into three hydrogels with the same composition before and after being suspended in the solution for an hour. Building on the work of Wereszczynska and Szczesniak \cite{wereszczynska_mri_2021}, a gelatin hydrogel was chosen that approximately simulates the diffusion properties of human muscle. We mix the gelatin in 1x Phosphate Buffered Solution (PBS) instead of deionized water to slightly increase the contrast between the hydrogel and the ternary solution. Initially, near zero volume fractions of DMSO and glycerol are present in the hydrogel. Each scan takes 15~min, so some DMSO and glycerol will have diffused into the gel during the first scan. After an hour, DMSO and glycerol volume fractions increase. The hydrogel volume fraction remains mostly constant as expected. The water volume fraction in the gel decreases as DMSO and glycerol occupy more of the volume. Furthermore, the results show that DMSO and glycerol do not diffuse at equal rates.

\begin{figure}
    \centering
    \includegraphics[width=0.97\textwidth]{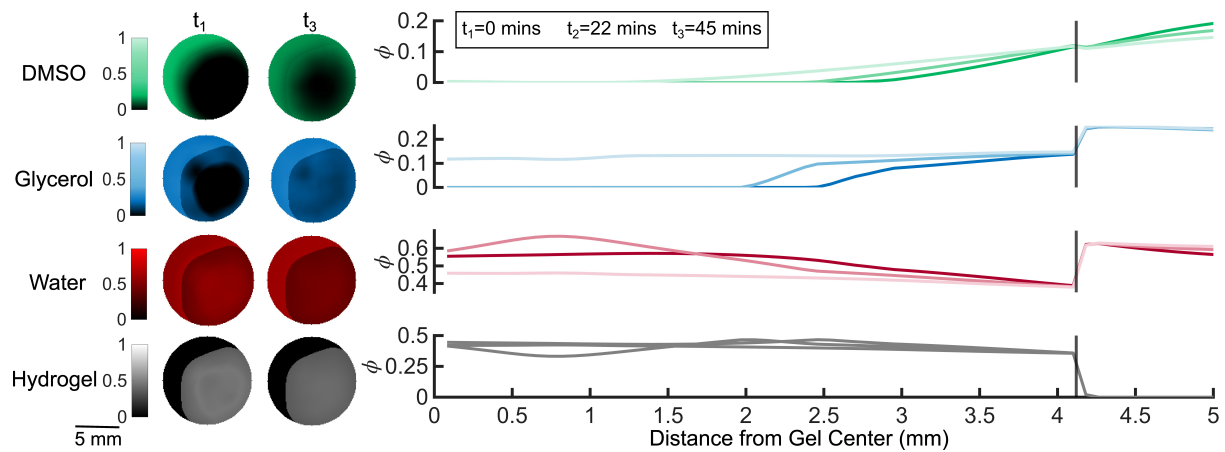}
    \caption{The average radial change in volume fractions inside and outside the gel. DMSO and glycerol are seen to increase within the gel over time whereas the water decreases. Lighter shades are at later times. The gel edge is denoted with a vertical black line. The volume fraction of the hydrogel is measured to be constant, as expected, since the hydrogel is not being transported.}
    \label{fig:timeresolve}
\end{figure}

Figure \ref*{fig:timeresolve} is a quantitative comparison of the diffusion of DMSO, glycerol, and water over time, demonstrating the ability to gather spatiotemporal data. Decomposition results are shown for three time points: immediately after depositing the gel into the solution, 22~min after deposition, and 45~min after deposition. The average radial profile from the gel center shows how the distribution of each component changes over time. Glycerol is observed to diffuse into the hydrogel faster than DMSO. Images show how the distribution within the hydrogel changes at the corresponding times. Although not explored in this study, temporal measurements such as those in figure \ref*{fig:timeresolve} can potentially be used to measure the diffusion coefficient matrix for the CPA components -- a critical parameter for CPA perfusion modeling.

\begin{figure}
    \centering
    \includegraphics[width=1\textwidth]{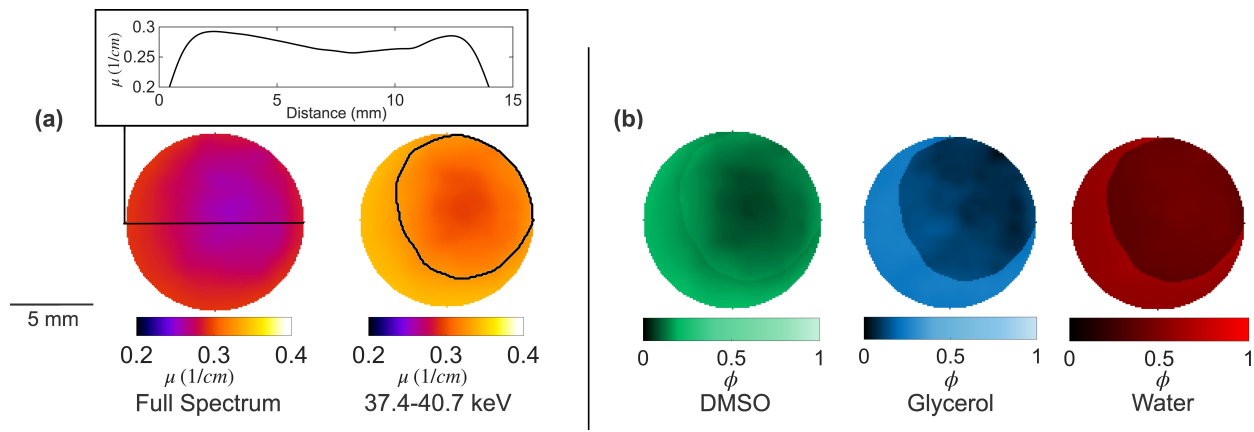}
    \caption{(a) Left: The full spectrum reconstruction of the hydrogel in solution after 90~min. The image shows a smoothly decreasing image intensity towards the inside of the hydrogel. With conventional energy-integrating X-ray CT this might suggest the CPA components are distributed in the same manner. Right: The reconstruction using the 37.4 -- 40.7~keV photon energy bin. The hydrogel has sufficient contrast to be visible; the outline of the segmentation is shown. (b) Images of the same hydrogel and time point based showing the decomposed volume fractions. The decomposed volume fractions show that the components are not distributed in the same manner as might be expected from (a).}
    \label{fig:recon}
\end{figure}

It is helpful to compare MSCT images to conventional X-ray CT images commonly used in cryopreservation literature. To do so we reconstructed the data after integrating the photon counts across the full photon energy spectrum to make energy-integrated projection images, assuming that the detection efficiency for the PCD and an energy-integrating detector are equal. Figure \ref*{fig:recon}a shows a full-spectrum reconstruction of the hydrogel after it has been suspended in the solution for 90~min, emulating a typical X-ray CT measurement that is not photon energy-resolved. The hydrogel appears to have a smoothly decreasing distribution of image intensity towards the interior of the hydrogel. At first this might suggest the same smoothly decreasing distribution of the CPA components. Figure \ref*{fig:recon}b shows the volume fraction of each component in the same hydrogel at the same point in time. The components are not in fact roughly distributed as might be suggested by figure \ref*{fig:recon}a, highlighting how MSCT can improve upon conventional energy-integrated X-ray CT methods.

\begin{figure}
    \centering
    \includegraphics[width=0.8\textwidth]{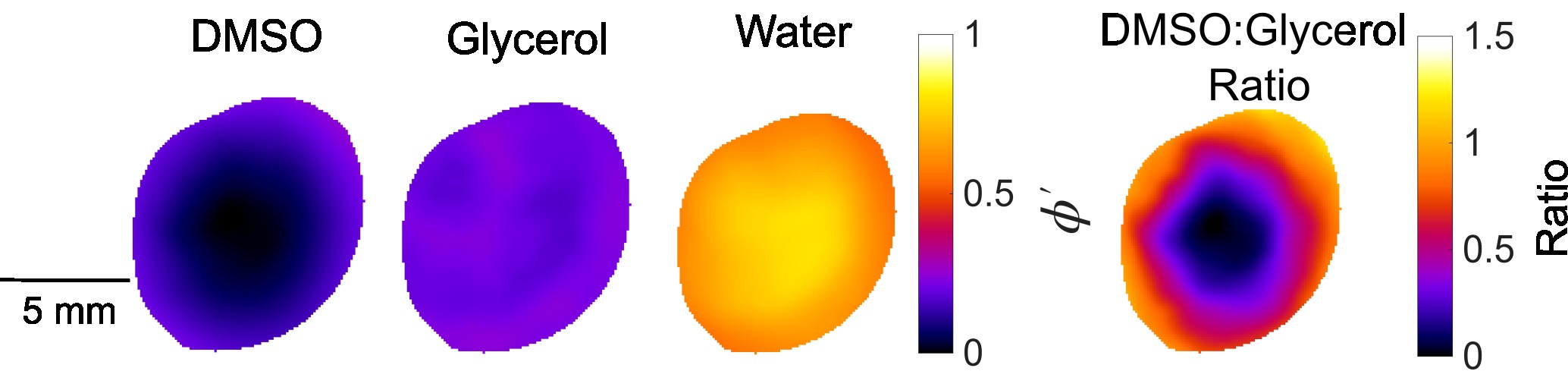}
    \caption{The solutal volume fraction, $\phi'$, shown 45~min. after depositing the hydrogel. The solutes are heterogeneously distributed. On the right, for example, is the ratio of DMSO volume fraction to glycerol volume fraction. Heterogeneous distribution of the solutal volume fractions means the combined solutal mass attenuation coefficient is non-constant.
    }
    \label{fig:solutalvolfrac}
\end{figure}

Figures \ref*{fig:geldecomp} and \ref*{fig:timeresolve} show that even an hour after the hydrogel is inserted into the CPA solution, the distribution of glycerol and DMSO remains different. That is, the two components do not diffuse homogeneously. We can calculate the solutal volume fractions from the measured component volume fractions to determine if the solutes diffuse homogeneously. Figure \ref*{fig:solutalvolfrac} shows that the heterogeneous diffusion means the solutal volume fractions change in space, which could introduce error to diffusion measurements taken with conventional X-ray CT via equation \ref*{eq:constatten}. A significant benefit of MSCT is that it does not require assuming homogeneous diffusion as it decomposes each of the components, and hence does not require that equation \ref*{eq:constatten} hold.

\section*{Discussion}

To show the utility of MSCT, we have validated and demonstrated MSCT measurements of CPA diffusion into a tissue-mimicking hydrogel phantom. The experiments in this study are enabled by an energy bin selection algorithm that balances detected photon flux and CPA component X-ray attenuation. The energy bin selection algorithm is designed to be easily implemented in a laboratory setting without requiring detailed X-ray system characterization. Prior work on energy bin optimization relies on computational, model-based approaches \cite{hu_optimizing_2025, nishigami_optimization_2025} that use detailed X-ray system characterization of sample and system scatter, fluorescence, and detection efficiency. Here, we achieve satisfactory results with a spectral projection image and empirical analysis. The measurements would not be feasible if bins are selected with other strategies, such as maximum flux and maximum contrast, which at first may seem intuitive. Even with more aggressive regularization to a lower condition number, the maximum flux and maximum contrast strategies cannot achieve the same performance for all components as the bins selected with the energy bin selection algorithm.

The bin selection algorithm is simplest to implement with PCDs that have many energy bins. The particular detector we use, the X-Card ME3, can use up to 128 energy bins with a minimum bin size of 1.1~keV. With the ME3 it is possible to image a broad spectrum of photon energies in a single exposure. Ostensibly 128 materials (one per bin) can be decomposed in a single CT scan with the ME3. Although, in reality much fewer are possible due to noise, energy bin size, and X-ray attenuation contrast limitations. Detectors with fewer energy bins would need to span the spectrum of photon energies with multiple exposures to use the energy bin selection algorithm developed in this study, though this is not a significant impediment.

Using an energy-resolving PCD has the added benefit of increasing the temporal resolution of the MSCT method. In dual-spectrum or some multi-spectrum imaging, the X-ray source, filter(s), or detector must switch settings between exposures in order to capture spectral information \cite{russo_handbook_2018,granton_implementation_2008}. Energy-resolving PCDs, on the other hand, can capture spectral information in one exposure. PCDs avoid challenges that arise if rapid diffusion occurs in between switching the source, detector, or filters. We take advantage of the energy-resolved PCD temporal resolution enhancements to capture scans in 15~min, which is fast enough to dynamically measure the diffusion of the CPA solution components, shown in figure \ref*{fig:timeresolve}. Faster measurement speeds are possible with a combination of a faster rotation stage, brighter X-ray source, and a higher count rate PCD.

For the CPA components used in this study, the goal is to maximize contrast between all of the components with one another. The balanced energy bin selection strategy maximizes contrast between all of the components instead of maximizing the total contrast magnitude, which can be dominated by one component. In other applications of the energy bin selection algorithm, however, the goal may be different. Alternative contrast metrics may in some cases be better than the metric used in this study. For example, in systems with components that have relatively balanced contrast across a range of photon energies, a better metric may be the L$_2$-norm of the attenuation coefficient contrast vector (which we use for the maximum contrast strategy).

The energy bin selection algorithm also requires an estimate of the detected spectrum as an input. We use the measured detected photon counts of one of the ternary solutions as an approximation of the expected detected spectrum, even when imaging both the second solution and the full system with the hydrogel. There is no guarantee that this approximation is valid in all cases. Other contrast metrics and better detected spectra inputs may be needed to achieve satisfactory results. Nevertheless, the bin selection algorithm is readily adaptable to different samples, metrics, and spectra. Improving empirical energy bin selection algorithms is an area for future research that expands the envelope of feasible MSCT experiments. Improved empirical methods for energy bin selection should be developed to include other energy-dependent phenomena such as multi-detection (in the case of PCDs) \cite{parker_experimentally_2022}, and detector photon energy point spread functions.

As PCD and laboratory X-ray source hardware improves, MSCT will become an increasingly capable method. PCDs are a relatively new technology that is still improving. New PCDs are expected to have better photon energy resolution, larger detection areas, and higher photon count rates that enable faster, more accurate MSCT decomposition. The ME3 detector used in this study is a line detector, so only one cross section of the system can be observed at a time. In the future, however, area detectors with more photon energy thresholds will become available that enable the full volume reconstruction of a sample. Furthermore, liquid metal jet (LMJ) X-ray sources can increase the detected flux by an order of magnitude over conventional tungsten-target X-ray sources, depending on the experiment \cite{parker_-lab_2024}. LMJ sources also use lighter target metals such as gallium and indium that have lower characteristic emission energies. Lower characteristic emission energies can be advantageous for biological applications as there is generally greater attenuation contrast between low molecular weight biological materials at lower photon energies. Radiation exposure must be considered in any medical application of MSCT, of course. Nevertheless, the higher flux at lower photon energies that LMJs can offer could take advantage of the improved contrast at those photon energies.

Figure \ref*{fig:timeresolve} shows that even with a more commonly used tungsten-target X-ray source we are able to distinguish between water, glycerol, and hydrogel -- three materials with relatively low X-ray attenuation contrast in the 10-100~keV photon energy range. Hydrogel phantoms are often used to mimic the diffusion properties of human tissue \cite{wereszczynska_mri_2021}. In these CPA diffusion experiments, we use an 11 w/w\% gelatin hydrogel. Figure \ref*{fig:timeresolve} shows that even as the volume fraction of the other components changes, the hydrogel volume fraction remains constant, as expected, giving us confidence in the hydrogel decomposition. This demonstrates the promise of MSCT to distinguish low contrast materials in biological systems without any contrast agents. Omitting contrast agents can be helpful since, if dissociated from the component(s) of interest, contrast agents may not diffuse at the same rate as those components. Measurement of contrast agent transport, therefore, is an imperfect measurement of true component transport. Furthermore, contrast agents that are tagged to the component(s) of interest may alter their diffusion properties, especially when a component might be transported across a cell membrane. Avoiding the use of contrast agents is one of the main advantages of the MSCT method introduced in this study.

In addition to sensitivity to low contrast components, distinguishing multiple components can be useful for assessing complete CPA loading. Calibrated uniform grayscale values have been used to investigate binary and multi-component CPA permeation and can yield valuable information when CPA loading is complete. As figure \ref*{fig:recon} shows, however, apparent uniform (single grayscale value) energy-integrated CT image intensity cannot guarantee uniform multi-component CPA distribution. If the mass attenuation coefficient is constant, then from equation \ref*{eq:ctdecomp} uniform image intensity implies uniform concentration. If, however, the mass attenuation coefficient of the combined solutes is not constant, then uniform image intensity does not necessarily imply uniform CPA concentration. One advantage of the MSCT method introduced here is that it removes the implicit assumption of constant solutal mass attenuation in standard X-ray CT measurements of CPA perfusion, and hence enables quantifying heterogeneous CPA component perfusion.

As we have demonstrated, the advantages of MSCT are particularly useful in vitrification perfusion applications. MSCT allows perfusion protocol development to be grounded in direct, quantitative measurements of component-resolved CPA distributions within a sample. In a vascular perfusion context, the same methodology could be used to image loading of a multi-component CPA cocktail into a kidney or liver via \textit{ex vivo} machine perfusion, revealing how different components penetrate, for instance, the renal cortex versus medulla over time and enabling targeted adjustment of perfusion pressures, temperatures, and concentration ramps to achieve organ-scale vitrifiability with minimal toxicity. The advent of vitrification for cryopreservation is predicted to enable cryogenic banking of transplantable organs. In this eventual scenario, MSCT-derived, component-resolved maps of CPA concentration could ultimately standardize assessments for determining whether an organ has been sufficiently equilibrated with cryoprotectants prior to cryo-banking and whether successful ice-free vitrification will be achieved.

MSCT composition measurements also have broad impact beyond cryopreservation. Multi-component fluid systems, for example, have recently attracted interest for the critical role they play in respiratory disease transmission and industrial processes \cite{lohse_fundamental_2022, lohse_physicochemical_2020, bourouiba_fluid_2021, longest_review_2024}. MSCT is a promising technique to measure the evolving chemical composition of microenvironments inside respiratory or industrially-generated droplets, for example. Moreover, developing the MSCT method further can allow the quantification of parameters such as the diffusion coefficient matrix of a multi-component system. Any system that involves dynamic spatial change of three or more components, or even two components with no fluorescence edges, could be a potential use case for MSCT.

\section*{Methods}

\subsection*{X-ray Imaging Settings}
We operate the source at 120~kVp and 200~\textmu A. At these settings, the focal spot is 7~\textmu m according to the manufacturer's data. The ME3 detector is placed 1,367~mm from the X-ray source. Samples are placed 140~mm from the source. Exposures are taken three 0.1~sec exposures combined into one 0.3~sec exposure time. The ME3 has a pixel size of 0.8~mm; the geometric magnification is roughly 9.76$\times$ so the effective pixel size is 82~\textmu m. The focal blurring is approximately 68.3~\textmu, so it is smaller than the effective pixel size. We do not expect focal blurring effects at these settings.

\subsection*{Energy Bin Selection Algorithm}
To evaluate equation \ref*{eq:binopt}, NIST mass attenuation coefficients and standard temperature and pressure density are used to approximate each component's linear attenuation coefficient. A single spectral projection image of the CPA solution is taken to estimate the number of detected photons, $I(E_i)$, that can be expected in each energy bin. With the ME3 PCD used in this study, this is possible with a single exposure. For PCDs with too few thresholds to span the photon energy range, a sweep of the photon energy range with multiple exposures would be necessary.

\subsection*{Bin Width Selection}
Bins are ideally infinitesimally narrow and still have sufficient flux to capture an image. Unfortunately, this is only possible to approximate with monochromatic X-ray systems such as synchrotrons. When implementing laboratory photon-counting MSCT, we must use bins with finite width due to detector and flux limitations. The width of a bin must balance achieving sufficient photon flux with error introduced to the X-ray attenuation coefficient due to multiple photon energies being detected in each bin.

Similar to the energy bin selection algorithm, an empirical approach determines the bin width. A flux-weighted moving average for different prospective bin widths is applied to the NIST attenuation coefficients \cite{j_h_hubbell_x-ray_2004}. This qualitatively models the penalty of a larger bin by shifting the linear attenuation coefficient away from the NIST value. An accuracy threshold is set at within 1\% of the NIST value at each photon energy. For each photon energy the maximum bin that achieves less than 1\% error between the weighted average smoothed and true NIST attenuation coefficients is the energy bin width to be used for that photon energy. During the energy bin selection, if that photon energy is chosen as optimal then the bin width selected in the manner described here is used.

\subsection*{Solutions \& Hydrogel Preparation}

\subsubsection*{CPA Solutions}
The required masses of each component necessary to result in the desired volume fractions in a nominal total solution volume of 50~mL are calculated. The density of each component is measured by weighing a known volume to verify against known standard densities. Table \ref*{tab:solmix} shows the densities and masses of each component in the two solutions. We note that the measured densities are very close to standard values, hence, we approximate to standard densities of 1.1 g/cm$^{3}$, 1.26 g/cm$^{3}$ and 0.998 g/cm$^{3}$ for DMSO, glycerol and water, respectively, for calculating the masses required to mix solutions with our desired volume fractions.

\begin{table}
    \centering
    \begin{tabular}{cccccccc}
     & & & & \multicolumn{3}{c}{} \\
    \multirow{2}{*} {Component} & \multirow{2}{*} {$\rho$ (g/mL)} & \multicolumn{3}{c} {Solution 1} & \multicolumn{3}{c}{Solution 2} \\ 
    & & $m$ (g) & $v$ (mL) & $\phi$ & $m$ (g) & $v$ (mL) & $\phi$\\
    \hline
    \hline
    DMSO & 1.12 & 16.6 & 15.1 & 0.297 & 11.4 & 10.4 & 0.205 \\
    \hline
    Glycerol & 1.28 & 12.6 & 10.0 & 0.197 & 18.8 & 14.9 & 0.293 \\
    \hline
    Water & 0.999 & 25.6 & 25.7 & 0.506 & 25.4 & 25.5 & 0.502\\
    \hline
    \end{tabular}
    \caption{Two solutions are tested with different amounts of DMSO and glycerol to validate the decomposition method and energy bin selection algorithm. The measured density of each component is shown; they are close to the reference densities.}
    \label{tab:solmix}
\end{table}

\subsubsection*{Hydrogel}
The use of hydrogel phantoms to mimic the diffusion properties of human tissue is common for applications such as diffusion-weighted magnetic resonance imaging (DW-MRI) \cite{wereszczynska_mri_2021}. Hydrogels are a popular choice for phantoms because they are easy to fabricate and it is possible to tune their properties predictably \cite{hacker_criteria_2022}. Previously, Wereszczynska and Szcześniak \cite{wereszczynska_mri_2021} explored the diffusion coefficients of forty types of gelatin or agar based phantoms in order to mimic the diffusion properties of human cortical gray matter, cortical white matter, spleen, and muscle tissue. Based on their results, an 11 w/w\% gelatin hydrogel was chosen for this study to roughly mimic the diffusion coefficient of human muscle, which is 2.4$\times10^{-3}$~mm$^2$/s \cite{wereszczynska_mri_2021}. Furthermore, we used 1x PBS solution as the solvent for the hydrogel instead of DI water which is the ternary solution solvent in order to enhance the contrast between the hydrogel and the ternary solution. We still expect the phantom to roughly mimic the diffusion properties of human muscle tissue.

\subsection*{Multi-Spectral Reconstruction}

Except for the energy-integrated image comparison, images from each energy bin are reconstructed independently. For reconstruction, we use the Simultaneous Iterative Reconstruction Technique (SIRT) combined with Total Variation (TV) minimization. Other iterative reconstruction techniques that could be used include SART, maximum likelihood, and penalized likelihood. Comparing the performance of these reconstruction algorithms could be an interesting area for further study and might be necessary when applying MSCT to limited, noisy projection data. The combined SIRT-TV method is a practical choice in our case because we have a large projection set (300 per scan) and it provides superior reduction in noise and beam hardening artifacts compared to the analytical FBP approach. SIRT+TV also does not require ray-dependent modeling required by SART or the statistical modeling inherent to maximum-likelihood and penalized-likelihood approaches. Previous work on the use of traditional X-ray CT for concentration measurement in cryopreservation uses high source acceleration voltages to avoid beam hardening artifacts \cite{bischof_use_2007,corral_assessment_2015}. As our method combines reconstructions from multiple energies and uses SIRT+TV, we can significantly reduce beam hardening artifacts. Furthermore, SIRT can achieve comparable spatial resolution to FBP with reduced noise, especially after an adequate number of iterations, which makes it a powerful tool allowing the combination of low and high energy reconstructions for the concentration decomposition \cite{anam_investigation_2019}. 

The robustness of the SIRT algorithm allows incorporation of regularization and prior knowledge through, in our case, TV \cite{rasmussen_improved_2021}. Total variation minimization is used to incorporate the knowledge of smooth volume fraction gradients by penalizing large spatial gradients to the SIRT reconstruction framework. We apply total variation using the Chambolle algorithm \cite{chambolle_algorithm_2004} which is a dual optimization approach that combines data fidelity with gradient reduction to preserve edges and important reconstruction features while smoothing the reconstruction. The combined approach of SIRT and TV offers an advantage compared to FBP in accurately differentiating and quantifying materials in MSCT while effectively reducing noise and artifacts.

\section*{Data Availability}
The data that support the findings of this study are available from the corresponding authors upon request.

\section*{Code Availability}
Three code scripts are provided with this study. One python script shows the reconstruction code. Two MATLAB scripts are also provided which facilitate the volume fraction decomposition and the photon-energy bin algorithm.

%%% BIBLIOGRAPHY ITEMS %%%

\section*{Acknowledgments}
This research benefited from support of AFRI Competitive Grant no. 2020-67021-32855/project accession no. 1024262 from the USDA National Institute of Food and Agriculture (grant administered through AIFS: the AI Institute for Next Generation Food Systems. https://aifs.ucdavis.edu.) This work received financial support from the National Science Foundation (NSF) Engineering Research
Center for Advanced Technologies for Preservation of Biological Systems (ATP-Bio), NSF EEC \#1941543.

\section*{Author Contributions Statement}

A.A. performed data collection, processing, analysis method. J.P. conceived the study and MSCT analysis method, developed energy bin selection algorithm. A.A. and J.P. analyzed the data; main authors of the manuscript. A.N. and B.C. advised on cryopreservation, and CPA solutions; contributed to the manuscript. L.W. advised on hydrogels; contributed to the manuscript. B.R. provided laboratory resources for hydrogels and CPA solutions. S.M. conceived the MSCT method used; advised on data collection, processing, and analysis; provided funding and laboratory resources for the implementation of this study; edited the manuscript. All authors reviewed the manuscript.

\section*{Competing Interests Statement}
Alaa M. Ali, Jason T. Parker, Linnea Warburton, and Simo A. M\"akiharju have no competing interests to declare. Brooke S. Chang, Anthony N. Consiglio, and Boris Rubinsky have a financial stake in Biochoric Inc., a company working to commercialize various cryopreservation-related technologies.

\appendix
\newpage
%\documentclass[fleqn,10pt]{article}
%\usepackage[utf8]{inputenc}
%\usepackage[T1]{fontenc}
%
%%%% ADDED
%\usepackage[margin=1in]{geometry}
%\usepackage{hyperref}
%\usepackage{soul}
%\usepackage{subcaption}
%\usepackage{graphicx}%
%\usepackage{multirow}%
%\usepackage{array}
%\usepackage{amsmath,amssymb,amsfonts}%
%\usepackage{amsthm}%
%\usepackage{mathrsfs}%
%\usepackage[title]{appendix}%
%\usepackage{xcolor}%
%\usepackage{textcomp}%
%\usepackage{manyfoot}%
%\usepackage{booktabs}%
%\usepackage{algorithm}%
%\usepackage{algorithmicx}%
%\usepackage{algpseudocode}%
%\usepackage{listings}%
%\usepackage{rotating}%
%\usepackage{lineno}%
%\usepackage{accents}
%\newcommand{\todo}[1]{\textcolor{red}{TODO #1}}
%%%%%
%
\title{Supplemental Information – Time-Resolved Multi-Spectral X-ray Computed Tomography of Cryoprotectant Diffusion Into Biomimetic Material}

\maketitle
%\author{Alaa M. Ali$^+$,\\
%        \and
%        Jason T. Parker$^+$,\\
%        \and
%        Anthony N. Consiglio,\\
%        \and
%        Brooke S. Chang,\\
%        \and
%        Linnea Warburton,\\
%        \and
%        Boris Rubinsky,\\
%        \and
%        Simo A. M\"akiharju}
%\date{}

%\begin{document}

\flushbottom

\begin{center}
    Department of Mechanical Engineering, University of California, Berkeley, Berkeley, CA 94720, USA
    
    $^+$These authors contributed equally to this work.
\end{center}

\thispagestyle{empty}

\subsection*{Image Acquisition Settings}

The X-ray source in use is an YXLON FXE225.99 TwinHead. The source head can be swapped between a transmission and a directional head. It is capable of achieving a maximum of 225~kVp and a maximum power of 64~W and 320~W with the transmission and directional heads, respectively. We use the directinonal head in this study. It is worth noting that higher source settings are achieved at the expense of a larger focal spot as the point spread function's full width half max increases with higher power. The source anode has a tungsten target placed at an angle of 22.5\textdegree~ that deflects high speed electrons and emits photons towards the detector. The source for this study is set to 120~kVp and 200~\textmu A. Based on the manufacturer's data, the focal spot blur at these settings is approximately 7~\textmu m.

The photon counting detector is a linear array of two ME3 detector units connected together. The ME3 units provide an energy resolution of 128 energy thresholds (minimum 1.1~keV bin width) which allows all required energy information to be collected in a single scan. The two detectors provide a pixel wide FOV of length 256 pixels. The pixel pitch is 0.8~mm. The ME3 is a line detector, so the imaging area is 204.8~mm across by 0.8~mm high. The source-to-detector and source-to-object distances are 1,367~mm and 140~mm respectively. This results in a geometric magnification of 9.76$\times$ and, thus, an effective FOV of approximately 20.48~mm $\times$ 0.08~mm and a nominal resolution of 82~\textmu m. The 128 thresholds range from 19.8~keV up to 160.6~keV increasing in 1.1~keV increments. The size of the focal spot blur in the detector plane after accounting for geometric magnification is 70~\textmu m which is less than the effective size of a single pixel. Hence, we do not expect notable focal spot blurring artifacts.

For the MSCT scans, we collect 9000 lines for every 360\textdegree~ continuous scan. A line is akin to a projection image. Each line is collected with an integration time of 0.1~s which is the maximum integration time per line possible using the ME3 detector. In order to increase the detected photon counts, we bin every three lines which results in a total of 3000 lines per 360\textdegree~ rotation for reconstruction. The effective exposure between lines increases to 0.3~s. The maximum counts per pixel per second never exceeds $9.32\times10^3$~counts/s/pixel which is safely below the detector's maximum count rate of $5.0 \times 10^6$~counts/s/pixel. Additionally, it is below the $2\times 10^6$~counts/s/pix threshold beyond which linearity losses increase to more than 14\%.

The rotation rate is set to 0.4\textdegree/s. Our sample is the 15mL Corning centrifuge tube (Part\# CLS 430766). Given the tube's diameter of approximately 17~mm, the CT scan resolution is 49~\textmu m which is smaller than a single pixel after magnification. Therefore we do not expect notable motion blur from rotation. Since the focal spot blurring and rotation motion blurring are both less than the effective pixel size, our resolution is limited by the effective pixel size.

\subsection*{CT Reconstruction Method}

The experimental procedure involves individual calibration scans of DMSO, glycerol, water and the hydrogel prior to submerging in the solution. Once the hydrogel is in the solution, a scan to capture the initial state is taken. Then, a continuous scan comprising three full rotations of the hydrogel in the solution is taken to dynamically image diffusion of each of the components. 

The forward projection matrix is constructed using the ASTRA Toolbox \cite{aarle_fast_2016} and the total variation minimization is done using Chambolle algorithm \cite{chambolle_algorithm_2004}. A total variation minimization step is implemented after every SIRT iteration. To prepare the data for reconstruction, we extract projections from the selected energy bins. The total number of photons within each energy bin is computed and normalized by the corresponding total flat field counts at those energies. We calculate the sinogram by taking the negative of the natural log of the normalized signal. The center of rotation is corrected for by shifting the projection geometry after looping over a range of pixels until the pixel shift at which the geometry aligns with the reconstruction is identified. The sinogram is then shifted by that pixel count to correct the center of rotation. Ring artifact correction is applied using Sarepy \cite{vo_superior_2018}. We use sarepy \texttt{remove\_all\_stripes} and the settings are shown in Table \ref*{tab:sarepy}. 

Reconstruction is done on a 256 $\times$ 256 pixel geometry which is the maximum possible resolution. The forward projection matrix is constructed using ASTRA toolbox \cite{van_aarle_astra_2015} and fed to an iterative reconstruction algorithm that updates the reconstruction over 300 iterations while performing 100 iterations of Chambolle iterative denoising or smoothing \cite{chambolle_algorithm_2004}. The SIRT learning rate is set to 0.5 and the TV minimization weight is set to 0.05. TV minimization preserves the edges and discontinuities in the image since it is a dual optimization that minimizes spatial gradients in the image while maintaining data fidelity. The main reconstruction parameters are summarized in Table \ref*{tab:ctrecon}.

\begin{table}
    \centering
    \begin{tabular}{rc}
    Parameter & Value \\
    \hline
    \hline
    Signal to Noise Ratio & 1.5\\
    \hline
    Median Filter Width (Large Stripes) & 8 pixels \\
    \hline
    Median Filter Width (Small Stripes) & 1 pixel \\
    \hline
    Drop Ratio & 0.1 \\
    \hline
    Dimension & 1 \\
    \hline
    \end{tabular}
    \caption{Sarepy ring artifact correction settings}
    \label{tab:sarepy}
\end{table}

\begin{table}
    \centering
    \begin{tabular}{rcc}
     & SIRT & TV \\
    \hline
    \hline
    Iterations & 300 & 100 \\
    \hline
    Weight & - &  0.05 \\
    \hline
    Learning Rate & 0.5 & - \\
    \hline
    \end{tabular}
    \caption{The main parameters used in the reconstruction step.}
    \label{tab:ctrecon}
\end{table}
\subsection*{Decomposition Method Background}

\subsubsection*{Energy-Integrating X-ray CT Decomposition}
Equation \ref*{eq:bllaw} shows the Beer-Lambert law as a function of photon energy, $\varepsilon$, and ray path $L$.
\begin{equation} \label{eq:bllaw}
    I(L, \varepsilon) = I_0(L, \varepsilon) Q(\varepsilon) \exp \left[ -\int_L \rho(\mathbf{x}) \varsigma(\mathbf{x}, \varepsilon) d\mathbf{x} \right]
\end{equation}
Here, $I(L, \varepsilon)$ is the image photon flux, $I_0(L,\varepsilon)$ is the source photon flux, 
$Q(\varepsilon)$ is the detector quantum efficiency, $\rho(\mathbf{x})$ is the density at location $\mathbf{x}$, and $\varsigma(\mathbf{x}, \varepsilon)$ is the mass attenuation coefficient. The source and detected photon flux are assumed to be constant over a single exposure, so $I(L, \varepsilon)$ is proportional to the detected photon counts.

\vspace{5mm}
\noindent \textbf{Note, for brevity and ease of reading we are using $\varsigma_i(\mathbf{x},\varepsilon)$ to represent the mass attenuation coefficient of material $i$ instead of the NIST notation: $\left( \frac{\mu}{\rho} \right)_i$} \cite{j_h_hubbell_x-ray_2004}. \textbf{The linear attenuation coefficient of $i$ is denoted $\mu_i$, as in the NIST notation.}
\vspace{5mm}

Energy integrating detectors cannot count photons or distinguish individual photons' energy. Instead, the image intensity is proportional to the deposited photon energy. In other words
\begin{equation}
    \tilde{I}(L) = \int_E \varepsilon I(L,\varepsilon) d\varepsilon.
\end{equation}
Assuming that the energy-integrated intensity approximately follows the Beer-Lambert equation with energy-integrated variables,
\begin{equation} \label{eq:enintBL}
    \tilde{I}(L) = \tilde{I_0}(L) \tilde{Q} \exp \left[ -\int_L \rho(\mathbf{x}) \tilde{\varsigma}(\mathbf{x}) d\mathbf{x} \right],
\end{equation}
where $\tilde{ \left[\cdot \right] }$ indicates the variable such that equation \ref*{eq:enintBL} is satisfied. These are referred to as the energy integrated variables. Note that these variables may not be equal to the energy integral of said variable.

Rearranging equation \ref*{eq:enintBL},
\begin{equation} \label{eq:sinogram}
    -\ln \left[ \frac{\tilde{I}(L)}{\tilde{I_0}(L)\tilde{Q}} \right] = \int_L \rho(\mathbf{x}) \tilde{\varsigma}(\mathbf{x}) d\mathbf{x}
\end{equation}
and performing reconstruction (inverse Radon transform),
\begin{equation} \label{eq:enintrecon}
    \tilde{R}(\mathbf{x}) = \rho(\mathbf{x}) \tilde{\varsigma}(\mathbf{x}) \equiv \tilde{\mu}(\mathbf{x}),
\end{equation}
where $\rho(\mathbf{x})$ is the total material density at location $\mathbf{x}$, $\tilde{R}$ is the reconstruction intensity, and $\tilde{\mu}$ is the energy integrated linear attenuation coefficient. Note that the reconstruction intensity is equal to the energy integrated linear attenuation coefficient. Assuming that the energy-integrated apparent mass attenuation coefficient, $\tilde{\varsigma}$, decomposes in the standard way for $K$ components \cite{j_h_hubbell_x-ray_2004},
\begin{equation}\label{eq:attendecomp}
    \tilde{\varsigma}(\mathbf{x}) = \sum_{i=1}^K \tilde{\varsigma_i} \chi_i(\mathbf{x}).
\end{equation}
where $\chi_i$ is the mass fraction of component $i$. Substitution into equation \ref*{eq:enintrecon} yields
\begin{equation}\label{eq:enintdecomp}
    \tilde{R}(\mathbf{x}) = \sum_{i=1}^K \tilde{\varsigma}_i c_i(\mathbf{x})
\end{equation}
where $c_i(\mathbf{x})$ is the concentration of $i$ at location $\mathbf{x}$.

For a solution of two components ($K=2$) equation \ref*{eq:enintdecomp} can be formulated as a calibration:
\begin{equation}\label{eq:binarydecomp}
    \tilde{R}(\mathbf{x}) = \tilde{\varsigma}_1 c_1 + \tilde{\varsigma}_2 c_2(\mathbf{x}) = A c_2(\mathbf{x}) + B,
\end{equation}
where $[\cdot]_{1}$ is the solvent and $[\cdot]_{2}$ is the solute. Once calibrated to determine $A$ and $B$, the solute concentration, $c_2(\mathbf{x})$, can be calculated from the reconstruction image intensity, $\tilde{R}(\mathbf{x})$. $A$ and $B$ are expected to be constant since they are material properties of the solute and solvent, respectively. The mass attenuation coefficient should not change for different concentrations of solute.

This approach does not readily extend to solutions with multiple components. Let $[\cdot]_{V}$ denote the solvent and $[\cdot]_{U}$ denote the combined solutes. Following the same procedure as for equation \ref*{eq:binarydecomp}, we can split equation \ref*{eq:enintdecomp} into the solvent and the solutes. That is,
\begin{equation}\label{eq:eintmulti1}
    \tilde{R}(\mathbf{x}) = \sum_{i=1}^K \tilde{\varsigma}_i c_i(\mathbf{x}) = \sum_{i \in U} \tilde{\varsigma}_i c_i(\mathbf{x}) + \tilde{\varsigma}_V c_V
\end{equation}
Only one energy-integrated image is available in standard X-ray CT, so decomposing further would result in an underdetermined system of equations. This necessitates treating the combined solutes as one material. To do so, we define
\begin{equation}\label{eq:combsolatten}
    \tilde{\varsigma}_U(\mathbf{x}) c_U(\mathbf{x}) \equiv \sum_{i\in U} \tilde{\varsigma}_i c_i(\mathbf{x}).
\end{equation}
Here, $\tilde{\varsigma}_U$ is the combined solutal mass attenuation coefficient and $c_U$ is the combined solutes concentration. Using the same decomposition from equation \ref*{eq:attendecomp} the solutal mass attenuation coefficient is given as
\begin{equation}\label{eq:solutalmassatten}
    \tilde{\varsigma}_U(\mathbf{x}) = \sum_{i \in U} \tilde{\varsigma}_i \chi'_i(\mathbf{x}),
\end{equation}
where $\chi'_i(\mathbf{x}) \equiv m_i / m_U$ is the mass fraction with respect to the solutes only. Substituting equation \ref*{eq:combsolatten} into equation \ref*{eq:eintmulti1} we get
\begin{equation}\label{eq:multicompCT}
    \tilde{R}(\mathbf{x}) = \tilde{\varsigma}_U(\mathbf{x}) c_{U}(\mathbf{x}) + \tilde{\varsigma}_V c_V(\mathbf{x})= A(\mathbf{x}) c_U(\mathbf{x}) + B
\end{equation}
The final equality is the form as is typically used for calibrations and measuring concentration with standard X-ray CT shown in equation \ref*{eq:binarydecomp}.

Combining solutes into one material in this manner requires a new condition be met compared to the binary solution. Calibrations are performed on uniform, well-mixed solutions of a given material with known concentration, so $A(\mathbf{x}) = \text{constant}$. In the multi-component case, however, $A(\mathbf{x})$ is a function of the solutal mass fractions at each location $\mathbf{x}$. From equations \ref*{eq:solutalmassatten} and \ref*{eq:multicompCT}:
\begin{equation}
    A(\mathbf{x}) = \tilde{\varsigma}_U(\mathbf{x}) \equiv \sum_{i \in U} \tilde{\varsigma}_i \chi'_i(\mathbf{x}).
\end{equation}
If the solutal mass fractions are not constant then $A$ is also not constant -- a contradiction with the calibration. This can be problematic because for a given solutal mass attenuation, $\tilde{\varsigma}_U$, equation \ref*{eq:solutalmassatten} is not satisfied by a unique set of solutal mass fractions. Different combinations of mass fractions can result in the same mass attenuation. In that case, it is impossible to know how much of each component is at a given location.

As a result, one \textit{must} assume that the components' mass fraction at a given location do not change relative to one another in order to obtain a result from equation \ref*{eq:multicompCT}. This implicit assumption is not commonly discussed in standard X-ray CT measurements of multi-component concentration and as we show in the main article may be frequently violated.

\subsubsection*{Multi-Energy CT Decomposition}
Similar to energy-integrating X-ray CT, we first use the Beer-Lambert law to model the image intensity. It is important to note that the Beer-Lambert law does not consider scatter and fluorescence effects. Both of these effects would be expected to affect the measurement, particularly at lower photon energies. Owing to the highly nonlinear and experiment-specific nature of these effects, we do not consider them in this model. 

Equation \ref*{eq:bllaw} can be written
\begin{equation}\label{eq:radonform}
    -\ln \left[ \frac{I(L, \varepsilon)}{I_0(L,\varepsilon)Q(\varepsilon)} \right] = \int_L \rho(\mathbf{x}) \varsigma(\mathbf{x}, \varepsilon) d\mathbf{x}.
\end{equation}
We calculate the left hand side of equation \ref*{eq:radonform} by capturing images with the object both in and out of the field of view -- the latter to measure the source intensity. Note that these variables are not energy-integrated because the PCD is energy-resolving.

The CT reconstruction operation in essence performs an inverse Radon transform on equation \ref*{eq:radonform}, which solves for the integrand. That is, the reconstruction $R(\mathbf{x}, \varepsilon)$ is given by
\begin{equation}\label{eq:recon}
    R(\mathbf{x}, \varepsilon) = \rho(\mathbf{x}) \varsigma(\mathbf{x}, \varepsilon).
\end{equation}
As above, the total mass attenuation coefficient for the mixed solution is given by
\begin{equation}\label{eq:effmu}
    \varsigma(\mathbf{x}, \varepsilon) = \sum_{j=1}^K \varsigma_j(\varepsilon) \chi_j(\mathbf{x})
\end{equation}
 for $K$ components, where $\varsigma_j(\mathbf{x}, \varepsilon)$ is the mass attenuation coefficient of component $j$ \cite{j_h_hubbell_x-ray_2004}, and $\chi_j(\mathbf{x})$ is the mass fraction of component $j$ \cite{j_h_hubbell_x-ray_2004}.

Substituting equation \ref*{eq:effmu} and $m = \rho v$ into equation \ref*{eq:recon}, we find
\begin{equation}\label{eq:conceq}
    R(\mathbf{x}, \varepsilon) = \sum_{j=1}^K \varsigma_j(\varepsilon) \Bigl[ \frac{m_j}{v}\Bigr](\mathbf{x}) = \sum_{j=1}^K \varsigma_j(\varepsilon) c_j(\mathbf{x}).
\end{equation}
Equation \ref*{eq:conceq} serves as the basis for quantitatively measuring the concentration of each component of the solution. Provided that 1) the solution components are known \textit{a priori} and 2) at least one reconstruction for each of $K$ energy bins are obtained, a system of linear equations can be formed to solve for the concentration of each component at each location in the reconstruction.

In order to constrain the solution we manipulate equation \ref*{eq:conceq} to be in terms of component volume fraction. Noting that $m_i \equiv \rho_i v_i$ and that by definition the volume fractions must be positive, less than one, and sum to one,
\begin{gather}
    R(\mathbf{x}, E_i) = \sum_{j=1}^K \bigl( \rho \varsigma(E_i) \bigr)_j \Bigl[ \frac{v_j}{v} \Bigr](\mathbf{x}) = \sum_{j=1}^K \mu_{j}(E_i) \phi_j(\mathbf{x}), \label{eq:volfraceq}\\
    \phi_j(\mathbf{x}) = \Bigl[ \frac{v_j}{v} \Bigr](\mathbf{x}) \in [0,1] \label{eq:constrain1} \\
    \sum_{j=1}^K \phi_j(\mathbf{x}) = 1 \label{eq:constrain2}.
\end{gather}
Here, $\mu_{j}(E_i) \equiv \left( \rho \varsigma(E_i) \right)_j$ is the linear attenuation coefficient $j$.

It is important to point out that the density, $\rho$, used throughout this derivation is the \textit{solution} density. The component apparent density in solution is not necessarily equal to the component particle density. As a result, the solution volume may not be the sum of the component volumes. The linear attenuation is a function of density, and therefore also changes after dissolution. We measured the apparent density of the solution with a scale to be approximately the same as the combined particle density of the components, so we assume $\mathring{\rho}_j \approx \rho_j$ where $\mathring{\rho}_j$ is the component particle density. Equation \ref*{eq:volfraceq} is now
\begin{equation}\label{eq:volfracapprox}
    R(\mathbf{x}, E_j) = \mathring{\mu}_{ij} \phi_j(\mathbf{x}) = A_{ij} \phi_j(\mathbf{x})
\end{equation}
where $A_{ij} = \mathring{\mu}_{ij} = \bigl( \mathring{\rho} \varsigma(E_i) \bigr)_j$.

In theory, one could use the mass attenuation coefficients from NIST \cite{j_h_hubbell_x-ray_2004} and known material particle densities to solve equation \ref*{eq:volfracapprox}. However, practical issues such as image noise, finite energy resolution, etc. introduce variance to the real measurement, necessitating a calibration. The attenuation coefficient matrix is calibrated as follows:
\begin{enumerate}
    \item Capture CT scans in each of the $K$ energy bins of each pure solution component.
    \item Select a region of interest in the reconstructed image to calculate an average linear attenuation coefficient for a given energy bin, $\left< \bar{\mu}_j(E_i) \right>$.
    \item Construct the linear attenuation coefficient matrix, $A_{ij}=\left< \bar{\mu}_{j}(E_i) \right>$, as a function of material and photon energy.
\end{enumerate}

We solve equation \ref*{eq:volfracapprox} for the volume fraction by regularized least squares regression (MATLAB function \texttt{lsqlin}) with the constraints shown in equations \ref*{eq:constrain1} and \ref*{eq:constrain2}.

% Calculating solutes attenuation %
\subsubsection*{Calculating $\varsigma_U$ From Multi-Spectral CT}
We calculate $\varsigma_U$ from the measured volume fractions to show the heterogeneity of the combined solutes attenuation, which violates the implicit assumption made in standard X-ray CT that it the combined solutes attenuation is constant.

For any location $\mathbf{x}$ we can break the total solution volume down into the solute and solvent volumes,
\begin{equation}\label{eq:totalvol}
    v = v_U + v_V.
\end{equation}
The solute volume fraction is defined as
\begin{equation}
    \phi'_i \equiv \frac{v_i}{v_U} = \frac{v_i}{v-v_V},
\end{equation}
where the second equality comes from substituting from equation \ref*{eq:totalvol}. Inverting and rearranging,
\begin{equation}
    \frac{1}{\phi'_i} = \frac{v-v_V}{v_i} = \frac{1}{\phi_i} - \frac{v_V}{v_i}\frac{v}{v} = \frac{1}{\phi_i}\left(1-\phi_V\right).
\end{equation}
From the final equality, we now have the solutal volume fraction as a function of the measured component volume fractions:
\begin{equation}\label{eq:solutalvolfrac}
    \phi'_i = \frac{\phi_i}{1-\phi_V}.
\end{equation}
Then, using the definition of $\varsigma_U(\mathbf{x})$,
\begin{equation}
    \varsigma_U(\mathbf{x}) \equiv \sum_{i \in U} \varsigma_i \chi'_i(\mathbf{x}) = \sum_{i \in U} \varsigma_i \frac{\mathring{\rho}_i}{\rho_U(\mathbf{x})}\phi'_i(\mathbf{x}) = \sum_{i \in U} \frac{\mathring{\mu}_i}{\rho_U(\mathbf{x})} \phi'_i(\mathbf{x}).
\end{equation}
The same density approximation is used as above in equation \ref*{eq:volfracapprox}. The density of the solutes alone is
\begin{equation}
    \rho_U(\mathbf{x}) = \sum_{j \in U} \mathring{\rho}_j \phi'_j(\mathbf{x}).
\end{equation}
Combining, we have
\begin{equation}\label{eq:solutalsigma}
    \varsigma_U(\mathbf{x}) = \frac{ \sum_{i \in U} \mathring{\mu}_i \phi'_i(\mathbf{x}) }{ \sum_{j \in U} \mathring{\rho}_j \phi'_j(\mathbf{x}) }.
\end{equation}
as shown in the main article. The over-circles are dropped in the main article for clarity. Equation \ref*{eq:solutalsigma} shows that if the solutal volume fractions are not constant then in general $\varsigma_U(\mathbf{x})$ is not constant.

\subsubsection*{Regularized Least Squares Regression}

The least squares regression algorithm reformulates equation \ref*{eq:volfracapprox} to a cost function and iteratively minimizes it for each voxel in the reconstruction to arrive at the optimal volume fraction of each component subject to the constraints shown in equation \ref*{eq:constrain1} and \ref*{eq:constrain2}. The cost function for least squares minimization is shown in equation \ref*{eq:lstsqreg}.
\begin{equation} \label{eq:lstsqreg}
    \phi(\mathbf{x}) = \min_{\phi^*} ||\mathbf{A} \phi^*(\mathbf{x}) - \mathbf{R}(\mathbf{x})||^2_{2}
\end{equation}
where $\mathbf{A}$ is the calibration matrix containing the linear attenuation coefficients of each component $j$ at each energy $i$, $\phi$ is a vector of the volume fractions of the components and $\mathbf{R}$ is a vector containing the linear attenuations at a single point in the reconstruction for each energy $i$. Ideally, this formulation should be sufficient to solve for the volume fractions. However, due to similarity in the linear attenuation of the components and the presence of small singular values, the linear problem becomes ill-posed. Even using just NIST values, our matrix is singular. Table \ref*{tab:simatten} shows the calibration attenuations in matrix A which also result in a singular matrix.

\begin{table}
    \centering
    \begin{tabular}{c|c|c|c|c}
    Energy Range (keV) & DMSO & Glycerol & Water & Hydrogel\\
    \hline
    \hline
    59.4-64.9 & 0.307 & 0.245 & 0.204 & 0.208 \\
    \hline
    48.4-57.2 & 0.368 & 0.261 & 0.218 & 0.219 \\
    \hline
    42.9-46.2 & 0.472 & 0.283 & 0.241 & 0.242 \\
    \hline
    37.4-40.7 & 0.584 & 0.306 & 0.265 & 0.265 \\
    \hline
    \end{tabular}
    \caption{Linear attenuation coefficients measured experimentally (cm$^{-1}$).}
    \label{tab:simatten}
\end{table}

The singular values confirm the ill-posed nature of our system and suggest that small perturbations in the reconstruction attenuation are sufficient to induce large errors in the optimal volume fraction result. This confirms the need for regularization and we reformulate the cost function to the standard regularized least squares form shown in \cite{golub_tikhonov_1999} and equation \ref*{eq:reglstsqreg}
\begin{equation} \label{eq:reglstsqreg}
        \phi(\mathbf{x}) = \min_{\phi^*} \left[ \Vert\mathbf{A}\phi^*(\mathbf{x}) - \mathbf{R}(
        \mathbf{x})\Vert^2_{2} + \lambda \Vert\mathbf{I} \phi^*(\mathbf{x}) \Vert^2_{2} \right]
\end{equation}
where $\lambda$ is a positive scalar regularizing parameter and $\mathbf{I}$ is the identity matrix. Tikhonov regularization reduces the condition number ($\kappa_{0}$) of the calibration matrix and improves numerical stability of the least squares regression algorithm. Regularization improves the ill-posed calibration matrix by adding a scaled identity matrix to drive it away from singularity. As the condition number is reduced and the numerical stability of the least squares regression solution improves, small perturbations in the calibration attenuations do not lead to large changes in the optimal solution. The regularizing parameter $\lambda$ is chosen by Monte Carlo simulations executed following the below steps:
\begin{enumerate}
    \item Calculate the expected reconstruction values for a solution of known volume fractions of DMSO, glycerol and water using the measured calibration attenuations.
    \item Simulate 100, 1,000, 10,000 and 100,000 different noise contaminated values randomly sampled from a normal distribution of the expected reconstruction attenuations and a bias error of 0.007 $1/cm$. The bias error is approximated as the maximum difference between the expected attenuation value and the measured solution reconstruction attenuation. The expected attenuation value is computed using the measured calibration matrix and known volume fractions.
    \item Define a search range for $\lambda$ between $10^{-8}$ and $10^{2}$. Loop over the chosen range and augment the $\lambda$-scaled identity matrix to the calibration matrix. Augment the solution vector with 0 at those rows for each noisy value.
    \item Run the least squares regression and compute the Euclidean norm of the residual and the Euclidean norm of the error between the optimal volume fractions and the expected volume fractions.
\end{enumerate}

\begin{figure}
    \centering
    \begin{subfigure} [b] {0.45\textwidth}
        \centering
        \includegraphics[width=\textwidth]{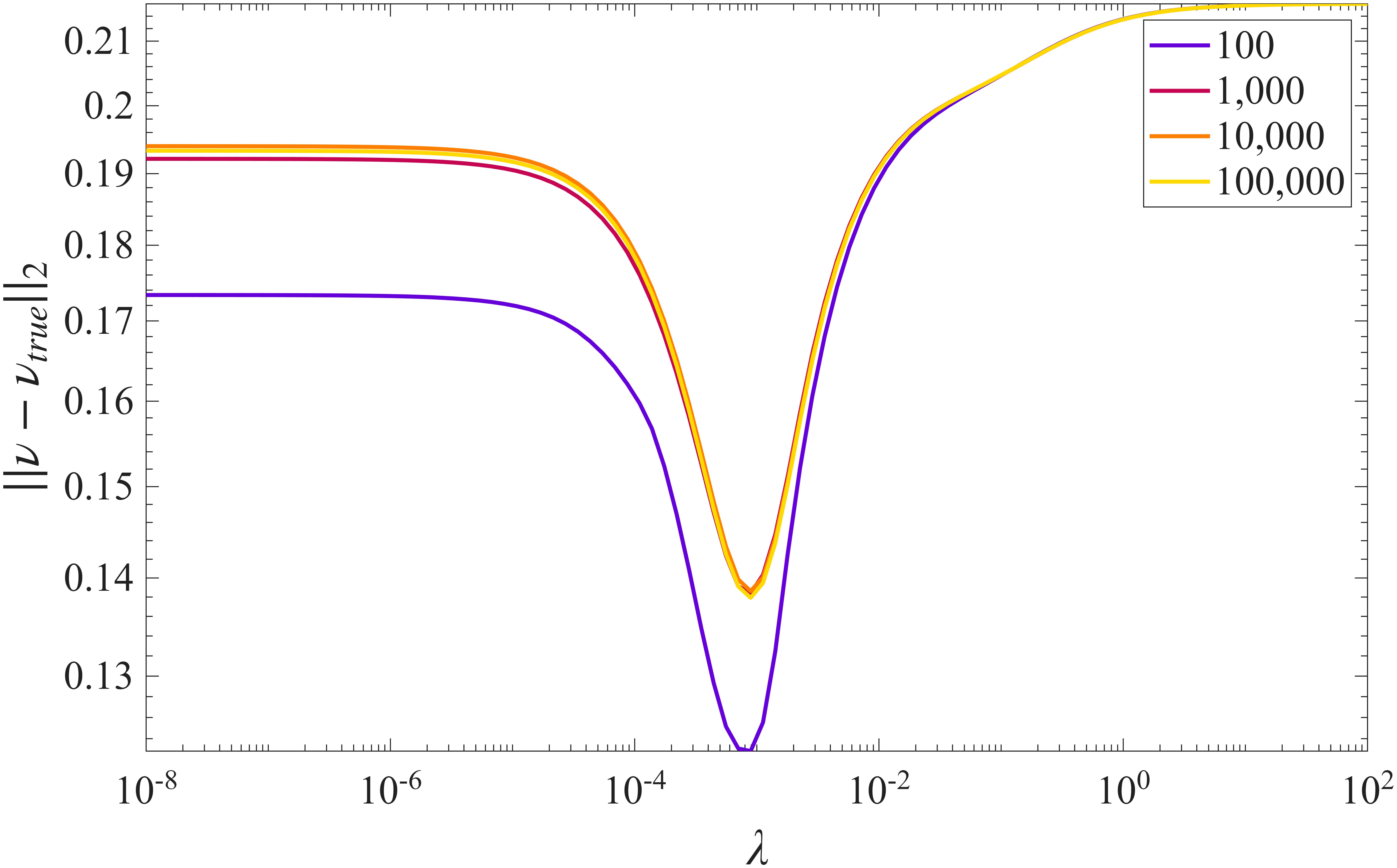}
    \end{subfigure}
        \begin{subfigure} [b] {0.45\textwidth}
        \centering
        \includegraphics[width=\textwidth]{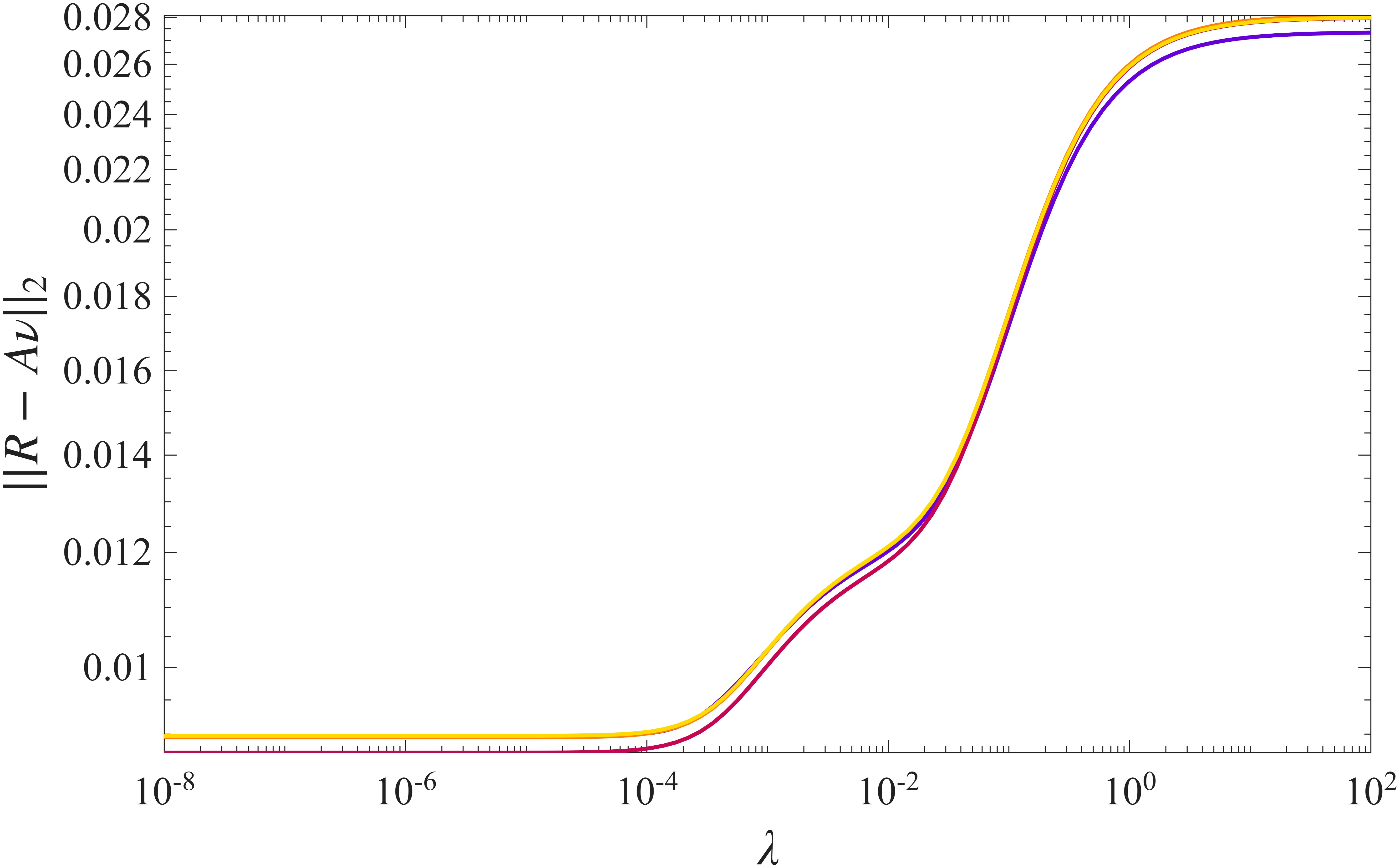}
    \end{subfigure}
    \caption{Residual norm and error norm  as the regularization parameter is changed for Monte Carlo simulations consisting of 100, 1,000, 10,000 and 100,000 noise contaminated attenuation values. The simulation is observed to converge. Noise level is randomly sampled from a normal distribution of the expected attenuation value of a solution of known volume fractions calculated using attenuations from calibration matrix and bias error of 0.007 chosen to approximate the reconstruction noise. The simulation converges to a minimum error at $\lambda = 8.51 \times 10^{-4}$. We find that $\lambda = 7.91 \times 10^{-4}$ works well for our images. We set that as our regularization parameter.}
    \label{fig:regularization}
\end{figure}

Figure \ref*{fig:regularization} shows the convergence of the Monte Carlo simulations and a minimum regularization parameter, $\lambda$. The minimum occurs at $\lambda = 8.51 \times 10^{-4}$. We find that $7.91 \times 10^{-4}$ works well with our images.

\subsubsection*{Measured Attenuation Accuracy}

\begin{figure}
    \centering
    \includegraphics[width=0.5\linewidth]{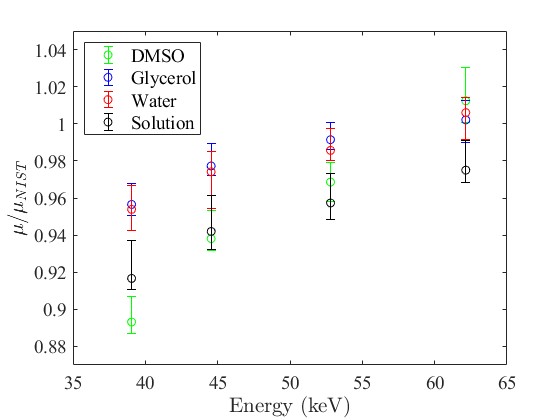}
    \caption{Ratio of measured attenuations to the true NIST attenuations. Error bars show the ratio of the minimum and maximum reconstructed values to the NIST attenuations. Errors are seen to be within about 10\% of the true values.}
    \label{fig:calibuncert}
\end{figure}

Figure \ref*{fig:calibuncert} shows the ratio of the measured attenuations to the true NIST attenuations and the error bars show the ratio of the minimum and maximum attenuation values to the true NIST values. The reconstruction of a homogeneous solution would be expected to yield a single attenuation. However, due to experimental limitations, we note that our error varies within roughly 10\% of the true values. We speculate that the heightened error at lower photon energies is due to scattered photons -- which tend to be lower energy -- and multi-detections where a photon's initial energy is divided into two or more detections of lower energy. A collimator may help reduce the effects of scatter. Using sources with lower photon energy may also help reduce scatter and multi-detections.

\subsection*{Energy Bin Selection Algorithm}

In an ideal world, the calibration returns the attenuation coefficient at a specific energy. In reality, though, PCDs have finite energy bins and real world imperfections that will disrupt the attenuation coefficient calibration matrix. Previous work in optimizing energy bins for multi-spectral CT \cite{nishigami_optimization_2025, brendel_empirical_2009, hu_optimizing_2025} has sought to minimize the effect that real-world imperfections have on the measurement. In this work, we use an empirical approach in order to develop a simple framework for practical in-lab use. In what follows we use ``optimal" to denote the results of our bin selection algorithm. We do not claim that the bin selection algorithm below selects the optimal energy bins of all possible sets. There may exist an algorithm which chooses a more optimal set.

A large attenuation coefficient contrast and small variance combine to create a robust calibration matrix. In other words, we are attempting to identify the set of optimal energy bins $\{E_O\}_{mn}$ between components $m$ and $n$ by
\begin{equation} \label{eq:mineq}
	\{E_O\}_{mn} = \max_{ \{E_i \}  } \left[ \frac{ \Delta_{mn}(E_i) }{ u_{mn}(E_i) } \right],
\end{equation}
where $\{E_i\}$ are a set of energy bins, $u_{mn}(E_i)$ is the relative uncertainty in $\Delta_{mn}(E_i)$ of materials $m$ and $n$ in energy bin $i$, with $\Delta_{mn}(E_i) \equiv \left| A_{im} - A_{in} \right|$. $A_{ij}$ is the attenuation coefficient as defined in the main article. The fraction term on the right hand side of equation \ref*{eq:mineq} is akin to a signal-to-noise ratio; the optimal bins can be thought of as those that maximize the signal to noise ratio.

Analytical forms for $u_{mn}(\varepsilon)$ and $\Delta_{mn}(\varepsilon)$ are difficult or impossible to obtain. The former requires a well-modeled imaging and CT reconstruction procedure; the latter requires fitting to NIST data or simplified models in limited photon energy regimes. To avoid these issues, the proposed metric will be tractable with empirical data.

% Image Quality Effects %
\subsubsection*{Image Quality Effects}
We know that the uncertainty in the reconstructed attenuation coefficient is partly driven by image noise, which in turn is driven by source flux. Exposure time, pixel size, etc. all affect image quality, although these parameters are energy-independent, and so can be neglected here.

The detector quantum efficiency is an energy-dependent parameter that will affect the image quality. Other effects, such as the energy point spread function and bin threshold width are likely important. This analysis is designed as a tool for quickly selecting energy bins in a laboratory setting, so we neglect these effects for simplicity. A more rigorous algorithm is an area for future research.

Lastly, attenuation by the components themselves will affect the image quality. With a low transmission percentage the image quality will be low without a long exposure time. To avoid introducing a model that requires detailed knowledge of the system to estimate the detected photon intensity, we simply capture a spectral projection image of the object. This method is practical in the laboratory which allows for quick experiment decision-making. Although this method does not account for any changes in the object composition with time or space, it provides a decent order of magnitude estimate for the detected photon intensity.

If there are $K$ components in the solution (including the solvent), then there are $S = \binom{K}{2}$ combinations of any two of those components. The relative uncertainty $u_{mn}(\varepsilon)$ is a set of vectors in $\mathbb{R}^S$ space. To simplify, we assume that the relative uncertainty has the same proportional relationship with the detected photon intensity, $I_{det}(\varepsilon)$, regardless of the materials chosen. That is,
\begin{equation} \label{eq:uprop}
	u_{mn}(E_i) \propto I_{det}^{q}(E_i).
\end{equation}

While $q$ is not known, it is a negative value. For a purely Poisson process, for example, the variance is equal to the mean. In that case, the relative uncertainty can be estimated as proportional to the inverse square root of the number of detected photons, i.e., $q = -1/2$. In this study, we take $q = -1$.

% Attenuation Contrast Effects %
\subsubsection*{Attenuation Contrast Effects}
If there is large contrast between the attenuation coefficients for different materials, then more uncertainty may be acceptable (although there is, of course, a limit to this). As with $u_{mn}(\varepsilon)$, $\Delta_{mn}(\varepsilon)$ is a set of vectors in $\mathbb{R}^S$ space. How these vectors are reduced to a list of scalars is a matter of choice. This reduction is defined as the contrast metric, $C(\varepsilon)$.

Below are two possible choices for the contrast metric:
\begin{align}
C(E_i) & = \left\Vert \Delta(E_i) \right\Vert_2 \label{eq:cont2norm} \\
C(E_i) & = \left\Vert \frac{  \Delta(E_i) }{ \left\Vert \Delta(E_i) \right\Vert_2 } - \mathbf{1}_S \right\Vert_2^{-1} \label{eq:contbalance}
\end{align}

Option 1 chooses energy bins with the maximum total contrast possible, even if this contrast is almost entirely due to two components at the expense of contrast with the others. Option 2 attempts to maximally balance the contrast between all of the components. $\mathbf{1}_S$ is the vector of balanced contrast between every combination of two components. For example, if two components have exceptional contrast in an energy bin, then with Option 1 that energy bin would be scored highly by the metric even if the other components cannot be distinguished from one another. Option 2 is used in this study to ensure that every component has the best chance of accurate volume fraction measurement. Alternative contrast metrics are possible, and some may perform better with other components. In this study, we use Option 2 since one of the components of our test solution, DMSO, has a much higher attenuation coefficient than the other two components, glycerol and water. We do not claim that this metric is the best of all possible choices.

Equation \ref*{eq:mineq} can now be written, as
\begin{equation} \label{eq:binmetric}
	\{E_O\} = \max_{ \{E_i \}  } \left[ \left\Vert \frac{  \Delta(E_i) }{ \left\Vert \Delta(E_i) \right\Vert_2 } - \mathbf{1}_S \right\Vert_2^{-1} \cdot I_{det}(E_i)\right] = \max_{ \{E_i \}  } \left[ P(E_i) \right].
\end{equation}

To ensure that the contrast metric and relative uncertainty metric are considered equally, each is normalized by its maximum value to range from 0 to 1. That is not shown in equation \ref*{eq:binmetric} for clarity.

Lastly, to incorporate the effects of finite bin widths, we take a moving average of the energy metric $P(E_i)$ with a kernel size equal to the bin width.
\begin{equation}\label{eq:binmetric_dsc}
    \{E_O\} = \max_{ \{ E_i \}  } \left[ \text{movmean}_{E_i}(P(E_i)) \right] = \max_{ \{ E_i \}  } \left[ M(E_i) \right].
\end{equation}
With $M(E_i)$ in hand, we choose energy bins in a cascade. First, the energy with the largest $M$ is chosen as the center of the first bin. The bin width at that energy is chosen as discussed below. All energies within that bin are removed from consideration to avoid bin overlap, which is not allowed by the detector. With those energies removed, the process is repeated until the number of desired bins has been produced.

\subsubsection*{Energy Bin Width}

An empirical approach is again used to determine the bin width. While in reality finite energy bin effects manifest in phenomena such as multi-detections \cite{parker_experimentally_2022} and energy point spread functions we use a simpler approach. Using a finite energy bin width is assumed to measure a photon attenuation that is equal to the moving flux-weighted mean of the NIST attenuation coefficient. A larger energy bin will spread the attenuation coefficient over more energies. By determining the difference between the moving flux-weighted mean of the attenuation coefficient and the NIST attenuation coefficient we can estimate an offset in the attenuation due to finite energy bin. Qualitatively, this captures the phenomenon of interest wherein regions of steep change in the attenuation coefficient with respect to photon energy are penalized with larger attenuation offset from the NIST values. A larger bin will, however, capture more photons for a higher quality image. Balancing attenuation offset with higher photon flux is what determines the bin width.

\begin{figure}
    \centering
    \includegraphics[width=0.5\linewidth]{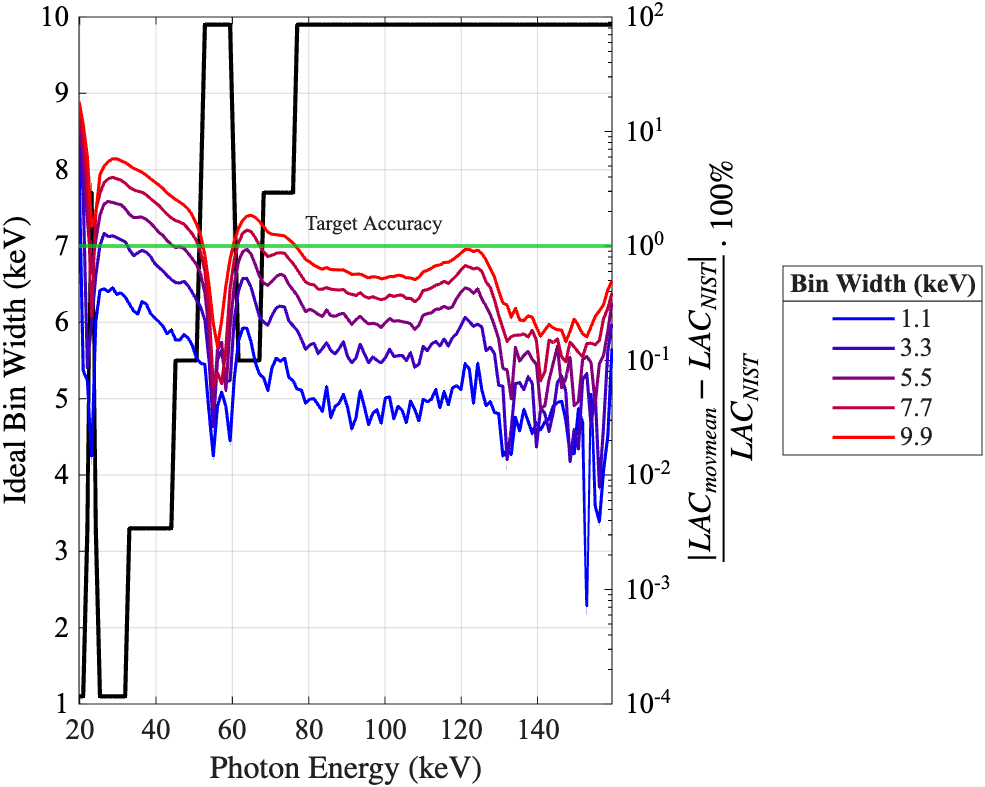}
    \caption{The linear attenuation coefficient (LAC) offset as a function of photon energy and bin width. The deviation from the true linear attenuation coefficient increases with increasing bin width. The largest bin width with an offset below the target accuracy (1\%) is chosen as the bin width to be used in the decomposition so as to maximize photon flux from a bigger bin without sacrificing decomposition quality.}
    \label{fig:binwidth}
\end{figure}

The offset is calculated as a percentage of the NIST value. For this study a 1\% offset is chosen as the maximum allowable offset. The chosen bin width is then the maximum width that achieves a 1\% offset or less. Figure \ref*{fig:binwidth} shows the attenuation offset as a function of photon energy and bin width. The offset increases with bin width and is more pronounced at lower photon energies where the X-ray attenuation of the solution is changing most.

\subsection*{Energy Bin Strategy Comparison}

Table \ref*{tab:condnum} shows the mean error for each selection strategy when regularization parameter or condition number of held equal. The balanced strategy developed above outperforms the other two strategies when both solution 1 and 2 are considered, regardless of whether regularization parameter or condition number are equal. The maximum flux strategy is comparable to the balanced strategy for solution 2, but the balanced strategy outperforms in solution 1.

\begin{sidewaystable}[!ht]
    \centering
    \begin{tabular}{cccccccccccc}
     & & & & \multicolumn{4}{c}{} \\
    \multirow{2}{*} {Method} & \multirow{2}{*} {$\kappa_0$} & \multirow{2}{*} {$\lambda \times 10^{-4}$} & \multirow{2}{*} {$\kappa$} & \multicolumn{4}{c} {Solution 1 Mean Error (\%v)} & \multicolumn{4}{c}{Solution 2 Mean Error (\%v)} \\ 
    & & & & DMSO & Glycerol & Water & Total & DMSO & Glycerol & Water & Total \\
    \hline
    \hline
    Balanced & 1069 & $7.91$ & 40.6 & 1.9 & 0.6 & 2.5 & 3.0 & 0.8 & 3.5 & 2.7 & 5.0\\
    \hline
    \multirow{2}{*}{Max. Flux} & \multirow{2}{*}{1253} & $7.91$ & 54.9 & 3.5 & 13.4 & 9.9 & 17.0 & 1.8 & 3.0 & 1.2 & 4.0\\
    &  & $16$ & 38.6 & 2.6 & 3.7 & 1.2 & 5.0 & 1.4 & 1.2 & 2.6 & 3.0\\
    \hline
    \multirow{2}{*}{Max. Contrast} & \multirow{2}{*}{354.3} & $7.91$ & 75.9 & 8.3 & 8.4 & 0.1 & 12.0 & 5.4 & 6.1 & 0.7 & 8.0 \\
    &  & $30$ & 39.6 & 7.3 & 6.5 & 13.8 & 17.0 & 4.7 & 4.0 & 8.7 & 11.0\\
    \hline
    \end{tabular}
    \caption{The mean error volume fraction (\%) for each component in the solution is lower with the optimal bins when either $\lambda$ or $\kappa$ is matched. Maximum flux and maximum contrast bins result in effectively useless measurements with errors of the same order of magnitude as the volume fractions. Choosing optimal energy bins is necessary for acquiring successful measurements in systems with components that have similar mass attenuation coefficients.}
    \label{tab:condnum}
\end{sidewaystable}

\section*{Hydrogels and CPA Solutions}

\subsection*{Hydrogel Preparation}
The hydrogel phantoms are fabricated using porcine gelatin with a 250 bloom value (\href{google.com/aclk?sa=L&pf=1&ai=DChsSEwj6w5vsn5iRAxVfIq0GHSxwIawYACICCAEQBxoCcHY&co=1&ase=2&gclid=Cj0KCQiA0KrJBhCOARIsAGIy9wCVa_HgbFMt6czXoAsXBYtq2iTSwFQ5SgRlfuMoinCVhc2t4jjkkssaAl-4EALw_wcB&cid=CAASZuRoKnYYqDDcNCp-Rxm2Bvdv8JwEfvosagUssxBfkS5qYDDWWVJM8HzJuK57dccmZUI3xu-Em6IEEUoPHHrfnP6SOWFAhQ_DImv7e6kPObSy0KltS7HN85n8qtryJcBjsedh7PZ8tA&cce=2&category=acrcp_v1_32&sig=AOD64_3FO3JpdHg2iszFPDXaFz8648EwlA&q&nis=6&ch=1&adurl=https://www.spectrumchemical.com/gelatin-250-bloom-powder-nf-ge105?gad_source%3D1%26gad_campaignid%3D21518051229%26gbraid%3D0AAAAAD_bqww0yOVFJhkN6OAX-NqbfBnwL%26gclid%3DCj0KCQiA0KrJBhCOARIsAGIy9wCVa_HgbFMt6czXoAsXBYtq2iTSwFQ5SgRlfuMoinCVhc2t4jjkkssaAl-4EALw_wcB&ved=2ahUKEwisnZTsn5iRAxVoOTQIHfR7FxUQ0Qx6BAgNEAE}{Spectrum Chemical}). 11 w/w\% gelatin was mixed into PBS, which had been chilled to 4~\textdegree C. The gelatin powder was added slowly and mixed manually with a spoon. Once the mixture appeared to be homogeneous, it was scooped into cylindrical molds and left to set in a 4~\textdegree C fridge overnight. The plastic molds were then peeled off, resulting in cylindrical gelatin phantoms with diameters of approximately 8~mm. The hydrogel phantoms were stored in a sealed container in a 4~\textdegree C fridge and used within 48 hours. The hydrogel was prepared by referring to the study conducted in \cite{wereszczynska_mri_2021}, however, using 1x PBS instead of water. The hydrogel is chosen to roughly mimic the diffusion coefficient of human muscle tissue.

\subsection*{CPA Solution Preparation}
The two CPA solutions are prepared at room temperature using the same DMSO, glycerol and deionized water containers. Masses of each component in each of the two solutions are calculated using standard densities and the volumes that would result in volume fractions of 0.3 DMSO, 0.2 glycerol and 0.5 water in the first solution and 0.2 DMSO, 0.3 glycerol and 0.5 water in the second solution. Table 2 in the main text summarizes the calculated masses. DMSO is first added into a volumetric flask followed by glycerol and lastly water. The solution was initially mixed by placing a stopper on top and slowly flipping the volumetric flask multiple times. Lastly, the solution was transferred to a 50 mL container and stir bar was used to mix the solution continuously for approximately 15 minutes.

%%% BIBLIOGRAPHY ITEMS %%%
%\bibliographystyle{unsrt}
%\bibliography{pcd_bibliography}

%\end{document}

\end{document}